\documentclass[english,11pt,aps,prd,eqsecnum]{revtex4}
\usepackage{amssymb,amsmath,amsthm,graphicx,amscd}
\usepackage{enumerate,color,comment,ulem,bm}
\usepackage[mathscr]{eucal}
\usepackage{xcolor}
\usepackage[hidelinks]{hyperref}

\usepackage{orcidlink}

\makeatletter
\renewcommand*\env@matrix[1][\arraystretch]{%
  \edef\arraystretch{#1}%
  \hskip -\arraycolsep
  \let\@ifnextchar\new@ifnextchar
  \array{*\c@MaxMatrixCols c}}
\makeatother

\begin{document}

\title{Atom-Field-Medium Interactions I: Graded Influence Actions for \\ $N$ Harmonic Atoms in a Dielectric-Altered Quantum Field}
\author{Jen-Tsung Hsiang\orcidlink{0000-0002-9801-208X}}
\email{cosmology@gmail.com}
\affiliation{College of Electrical Engineering and Computer Science, National Taiwan University of Science and Technology, Taipei City, Taiwan 106, R.O.C.}
\author{Bei-Lok Hu\orcidlink{0000-0003-2489-9914}}
\email{blhu@umd.edu}
\affiliation{Maryland Center for Fundamental Physics and Joint Quantum Institute,  University of Maryland, College Park, Maryland 20742, USA}
\date{August 5, 2024}

\begin{abstract} 
This series of papers has two broader aims: 1) Construct a theory for multi-partite open quantum systems comprising several layers of structure with self-consistent back-actions. Develop the graded influence action formalism \cite{BehHu10,BH11} to account for the influences of successive sub-layers on the dynamics of the variables of interest. 2) Apply these methods to the study of atom-field-medium interactions and highlight their merits over conventional methods. We consider a system of $N$ harmonic oscillators, modeling the internal degrees of freedom (idf) of $N$ neutral atoms (A), interacting with a quantum field (F), scalar here, for simplicity,  altered by the presence of a dielectric medium (M). In this paper we use the coarse-grained and stochastic effective actions in the influence functional formalism to derive the stochastic equations for the reduced density matrices of the dynamical variables in the successive layers of structure. The word `graded' refers to the specific ordering of the coarse-graining procedures. Three layers of  coarse-graining are performed, firstly, integrating over the common bath of the dielectric oscillators results not only in the appearance of necessary dissipative properties of the dielectric but also essential nuanced features such as nonMarkovian spatial correlations in the dielectric.  Secondly, integrating over the medium variables as a whole results in a dielectric-modified quantum field, the influence of the medium on the quantum field manifesting through a frequency-dependent permittivity function. Finally, integrating over this dielectric-altered quantum field which interacts with the idfs of the atoms yields an influence action. From it we obtain the stochastic equation of motion which describes the nonequilibrium stochastic dynamics of the idf of the atoms interacting with a dielectric medium-modified quantum field. In Paper II we proceed to calculate the nonequilibrium covariant matrix elements of the correlation functions of the idf of $N$-atoms in a dielectric-altered quantum field, which are useful for probing many basic quantum information issues, such as the entanglement dynamics in AFM interactions. 
\end{abstract}

\maketitle

\hypersetup{linktoc=all}
\baselineskip=18pt
\allowdisplaybreaks

\tableofcontents

\section{Introduction}
The goal of this series of papers is to establish a unified and self-consistent  theoretical framework to treat the interaction of moving atoms (or systems with internal degrees of freedom, imperfect mirror ) with a quantum field in the presence of a medium (a dielectric in an infinite half space, a conductor, a mirror). The demands we set forth for the this framework are that,   i) it is based on \textit{microscopic physics modeling}, invoking quantum field theory descriptions, rather than a semi-phenomenological description,  ii) the \textit{noise and fluctuations} in the different components are derivable from the make up of their own constituents,  e. g., from the fluctuations of their quantum field environments, \textit{not} put in by hand, and iii) it is capable of treating \textit{fully nonequilibrium dynamics} in real time \cite{CalHu08}; this means that it can reproduce results from, but not restricted to,  linear response theory which depends on the assumption of an equilibrium condition.  

We first mention the scope of this program, then the methodology, comparing ours with other major approaches, its merits weighed against its shortcomings, then describe briefly the main themes of this program, and our goals in this and subsequent papers of this series. 
  
{\it The Scope}. Examining pairwise the three major components of atom, field and medium,  1)  Atom-Field (A-F) Interaction has been the foundation of high profile and impactful fields such as quantum optics and  atom optics for at least half a century, from old  topics like Lamb shift and spontaneous emission  to more recent topics like laser cooling.  The newer issues of A-F interaction from 1995 onward entail cold atom physics, which has become a dynamic field by itself,  and fundamental issues of quantum information such as quantum decoherence and entanglement playing central roles in quantum computing, communication, control and metrology.  2) For  Mirror-Field (F-M) Interaction, we need only mention the famous Casimir effect \cite{Casimir,CasimirRev,CasimirRev1,CasimirRev2,CasimirRev3}, and for nonrelativistically moving mirrors \cite{Moore}, the dynamical Casimir effect \cite{DCErev};  3) For Atom-Mirror/dielectric (A-M) interaction, the Casimir-Polder effect \cite{CasPol,IntHenAnt} and for moving atom-dielectric interaction~\footnote{Both dynamical Casimir effect (DCE) and quantum friction (QFric) have cosmological analogs:  DCE corresponding to particle creation in an evolutionary universe \cite{Par69,Zel70}, QFric entails backaction of dynamically excited vacuum fluctuations, which is intrinsically a nonMarkovian process \cite{CalHu87}. Thus, `viscosity' would be a more appropriate word than  `friction' because the quantum fluctuation-induced force is a reactive force, like inductance rather than resistance in a RLC circuit analogy. See \cite{Hu82,XBH,DCE-CPC} for further discussions.},  quantum friction \cite{QFric,QFric1,QFric2,QFric3,QFric4,QFric5,QFric6,QFric7,QFric8,IntBehQFric,IntBehQFric1,IntBehQFric2,IntBehQFric3}.  Bringing all three components (A-F-M) together we enter into the fast expanding and rich field of quantum optomechanics \cite{QOM,QOM1,QOM2,QOM3,QOM4,QOM5,QOM6,QOM7,QOM8,QOM9} with many hybrid modeling schemes in atom and optical physics involving matter media (cavities, traps, moving mirrors, vibrating membranes, down to noise in the coatings in LIGO detectors, etc.).

\subsection{Key issues and existing popular approaches}
In classical electrodynamics  the role of the medium is played by a dielectric polarization tensor and its function described by the susceptibility function in response theory. Conceptually, for a linear (temporal-) dispersive material, the Kramers-Kr\"onig relations imply its accompanying dissipative nature. When one applies quantum electrodynamics to a dissipative medium: there are three basic issues to reckon with: How to 1) preserve the canonical relation in field quantization 2) explain how dissipation enters and 3) identify the source of noise from microscopic physics.

\subsubsection{Field quantization and quantum states}
Naive implementation of field quantization in the presence of such media often leads to inconsistency in the equal-time commutation relation, due to the decaying field amplitude. Several noticeable ways to deal with this issue have been presented:  In a \textit{microscopic modeling approach}, Huttner and Barnett~\cite{HB} (HB) quantized the combined F-M system.  See also  Eberlein and Zietal~\cite{EZ}, from which the reader can find the important references in the intervening twenty years. The formal advantage of this approach is that the quantum field theory of the \textit{closed system} can be mathematically sound and theoretically complete.  The physical disadvantage,  in the present context, is,  if one was interested in the dynamics of one or two of its components, be it  F or M,  while coarse-graining away the remaining component(s), the specific subsystem dynamics is not easily derivable from the formal structures of the closed system. To begin with, the quantum state of the diagonalized closed system is not that of the  field or the medium. If one wants to follow the fully nonequilibrium dynamics, one would encounter the added difficulty that the same well-defined vacuum for the quantum state of the atom or the field, something one usually takes for granted, may not  exist at all times in its evolutionary history. This is similar to the difficulty of defining a vacuum for quantum field theory in a  dynamical setting, such as in dynamical Casimir effect or particle creation in the early universe. Any quantity which changes with time (or frequency in Fourier space) such as the real time dynamical susceptibility function, which usually requires an equilibrium condition to be well defined,  will also be affected.

\subsubsection{Microphysical basis of noise sources}
Another popular approach is stochastic  (SED) \cite{Lifshitz, Lifshitz1,AntPit} or macroscopic electrodynamics (MED) \cite{Bu12,Bu13} where the noise sources in the stochastic equations for the reduced systems, such as the Langevin or Fokker-Planck equations, are designed in such a way that the commutation relations of the canonical variables are enforced.   MED involves the macroscopic Maxwell’s equations but with the addition of some  local bosonic sources. The sources are  fixed by the requirement that the equal-time commutation relations of the quantized EM fields should be preserved.  In this semi-phenomenological theory   the  polaritons formed by the polarization field of the medium and the electromagnetic (EM) field play a central role.   The susceptibility function relies on the input from experiments, required to satisfy several general properties: the Kramers-Kr\"onig relation, the reality condition and the reciprocal theorem, and as such, are well behaved and mimic experimental reality.  MED is restricted  to near-equilibrium or steady state configurations \cite{BHjpa}.

From the way it is constructed,  the obvious advantage  of this approach is that  the response functions of the dielectric medium are well suited to the specific material or composition used in an experiment. The shortcoming of this approach,  as in most phenomenological modeling, is its microphysics theoretical basis, such as the changing quantum states of the components and the physical origin of the noise sources  and the mechanisms by which they drive the field.

\subsection{The present approach}
We take a microphysics approach to this problem, so its base is aligned with that of HB and EZ models mentioned above. However, we do not attempt to quantize the whole A-F-M closed system, rather, we adopt the theories of open quantum systems  to calculate the reduced dynamics of the specific component(s), which become open systems under the influence of,  and modified by, the other components,  coarse-grained in successive steps in an orderly manner. There are three components in A-F-M, and counting both the internal (idf) and external (edf) degrees of freedoms (dof) of the atom (A), the modified field (F) plus the medium oscillators (M) and their common bath (B), there are five sets of dynamical variables in our problem (See Table 1 of \cite{BH11}).  The technique which enables this approach is the Feynman and Vernon (FV) \cite{FeyVer} influence functional formalism.  We call the ordered sequence of influence actions after each coarse-graining which accounts for the influences of the lower-lying level of  components,  the \textit{graded influence actions} (GIA). Taking the functional variation of  the graded influence action with respect to that particular variable of interest, one can obtain its stochastic equation of motion.  

The reduced density matrix of a particular physical variable after a certain number of rounds of coarse-graining will contain the coarse-grained information of all the subsystems in the sub-layers. {Here we work at the level of the influence functional, focusing on obtaining the GIAs of each layer of structure. The explicit expression of the reduced density operator of the $N$-atoms can be obtained by taking the functional derivative of the final closed-time-path effective action.}  We shall show how to obtain the GIA layer by layer, starting with the common bath of the dielectric medium which provides the damping mechanism, to the coarse-grained dielectric variables, leading to a medium-altered quantum field. Then by coarse-graining this field we arrive at the dynamics of the atom interacting with the quantum field now imbued with the influences of the dielectric and the noise of its bath. This path integral approach averts the quantization issue while producing fully viable and consistent quantum and stochastic results.  For Gaussian systems such as the ones under  study here, with bilinearly coupled systems of harmonic oscillators, the emergence of noise as a classical forcing term can be exactly defined  with the help of the Feynman-Vernon (FV) Gaussian identity, which avoids guessing at or putting by hand any noise which is likely to violate in unbeknown ways basic constraint conditions such as the fluctuation-dissipation relations. Taking the variational derivative of the influence action for a particular physical variable yields the Langevin equation for that variable with noise given via the Feynman-Vernon  identity. It is in these ways -- its first- principles approach, its clear systemics, its  self-consistency, together with the traceable microphysical origins of noises,  that we see this approach possessing many distinct advantages. We will show the key steps of the  schematics in the sections below.  

In terms of  architectural structure and contents,  our series continues  the work of Behunin and Hu (BH) who introduced the graded influence action formalism for the calculation of the fluctuation forces between two atoms \cite{BehHu10}, and that between an atom and a dielectric (Casimir-Polder)\cite{BH11}. We consider an extended model where the dielectric oscillators interact with a common bath field, which is more realistic than  each having its own bath.  Doing so enables one to see the nonMarkovian (spatially nonlocal) features of the noises in the bath, which is absent in the case where each dielectric oscillator interacts only with its private bath.  Programmatically the approach closest to ours is that by Lombardo, Mazzitelli and co-authors \cite{LomMazz,LomMazz1,LomMazz2}. For other functional integral approaches to AFM problems, see, e.g., \cite{Kadar,Lee,Lee1}.

In the first few papers of this series we will consider the cases when the atom sits still, but our later papers will treat moving atoms or mirrors for the investigation of dynamical Casimir, Casimir-Polder effects and quantum friction.  Improvements over the conventional approach in the treatment of the mirror-field interactions, where the coupling is by way of the  radiation pressure \cite{Law,BC,BC1}, was the primary motivation in the proposal of the mirror-oscillator-field (MOF) model in \cite{GBH,SLH} (See also \cite{WangUnr}).  There, like here for the atom,  the mirror's idf is modeled by an oscillator  which interacts with the quantum field. With this setup one can treat mirrors of finite transmissivity in a dynamical way. The same method can also be applied to treat an atom interchangeably \cite{LinMOF}. Thus structurally the MOF model involving only three dynamical variables, the idf and edf of the atom or mirror, together with  the quantum field, is a simpler example of the applications of the graded influence actions. Our present work can be considered as an extension of the MOF model in the treatment of a dielectric medium, where M refers now to medium, not just limited to mirrors, {and the O refers to the representation of the medium by a collection of oscillators,  not just a single one. If one turns the single oscillator in the MOF model to represent the idf of an atom, it would become our present problem with an added edf of the atom.}  The full display would involve five dynamical variables  (B, D, F, I, E): bath, dielectric, quantum field, idf and edf of the atom  (see Table 1 of \cite{BH11}), the first two grouped under medium (M), the last two under atom (A), in the atom-field-medium interaction.

 Our present approach utilizes the coarse-grained effective action \cite{cgea,cgea1} which encapsulates in a consistent way the influence and backaction of the medium on the field. Then we present the influence action governing the dynamics of the atoms and the corresponding Langevin equations.

\section{Modeling and Dynamics from an Ordered Sequence of Coarse-grainings}

\subsection{Modeling of the atoms, the field and the medium}

\subsubsection{The atoms}
Our system consists of $N$ neutral, polarizable atoms following given trajectories in an ambient quantum massless scalar field outside a linear dielectric medium. We use two dynamical variables to capture the activities of the neutral atom, which is the center of  our attention. Its external, or center of mass (CoM) degree of freedom (dof) $\bm{z}(t)$,  marks where the atom is located at time $t$  in a (1+3) Minkowski spacetime. Its  internal (or scalar charge) degrees of freedom $ \chi(t)$, modeled by a quantum harmonic oscillator, interacting with the ambient quantum field, the massless scalar field $\phi(x)$,  outside of,  but close to,   the material medium \footnote{A comment on how good it is to model a real-life multi-level atom by a quantum harmonic oscillator, both for atom outside of and for the atoms depicting the dielectric constituents. This approximation inevitably ignores the unequal spacing of energy levels and other finer structures of a real atom, but so do the  two level atom model which is highly successful in capturing many salient properties of atom-field interactions. In fact at sufficiently low temperatures (where the thermal excitation energy is smaller than the transition energy between levels), when most activities occur in the two lowest lying levels, the harmonic atom approaches the two-level atom.}.

\subsubsection{The field} 
We use a massless scalar (spin 0) field $\phi(x)$ ($x$ is a simplified notation for the spacetime point with coordinate $x^\mu = (t,x^i)$, $i =  1$, 2, 3) to mimic the electromagnetic (spin 1) field which interacts with the idf of the atom, thus sparing the burden of carrying the polarization index. This is fine, since the two polarizations of an EM field  each can be represented by a massless scalar field. However this cannot capture the orientation or anisotropicity of the atomic dipole moment outside the dielectric.  As long as the physical effects of interest   do not depend sensitively on this factor we can take advantage of this simplification. The type of coupling between the atom's idf and the field assumed here is the common one used in quantum Brownian motion \cite{CalLeg,GS88,HPZ92,CRV}, so it is analogous to $\bm{d}\cdot\bm{E}$ in the case of EM interaction. We only consider the dipole moment and ignore any higher moments. The dipole moment of the neutral atom is modeled by a quantum harmonic oscillator. It is sufficient to describe small polarization, in particular when the charge distribution inside the atom does not deform too much. The  merit in making this approximation is its Gaussian nature, which grants tremendous simplification in calculations and allows for exact and analytic results. This is particularly convenient for the investigation of atom-field interaction of finite coupling strength.

\subsubsection{The medium}
The bulk linear dielectric medium of arbitrary shape is described by a three-dimensional lattice of non-interacting atoms, whose internal degrees of freedom are also modeled by the harmonic oscillators or an equivalent quantum field.

{A famous early  model was that of Fano \cite{Fano}, who proved that the Hamiltonian of long-wave excitations of matter is equivalent to that of an assembly of oscillators under very general assumptions. A later oeuvre of significance is that of Welsch and co-authors \cite{GruWel,DunWel,RaaWel}, where a quantization scheme for the full electromagnetic field in an inhomogeneous three-dimensional, dispersive, and absorbing dielectric medium was developed, and a unified approach to macroscopic QED in arbitrary linearly responding media was established, which contributed to a fuller development of macroscopic QED represented by the work of Scheel and Buhmann \cite{SchBuh,Bu12,Bu13} mentioned earlier.  Amongst the more recent work of relevance we mention \cite{Drezet,HanBuh,ForMia,Cia} as they touch  on the relation between the Hamiltonian operator approach, as used in the HB model, and the so-called Langevin noise formalism related to macroscopic QED, both based on canonical quantization.  We shall also address this important issue in this and the following paper, but from the open quantum system conceptual framework by way of the influence functional integral formalism. }

This model is suitable for the description of a simple insulator, regardless of any band structures. This implies that the wavefunction of the medium atom's excited state does not overlap with that of another atom in the medium. This ensures the orthogonality or independence among the atoms. The dimension of the dielectric body is supposed to be much greater than the scale of our interest,  such that the extension of the dielectric, when viewed by the atom outside the medium, is sufficiently large that the edge effects are assumed negligible.

In \cite{BH11} the idf $Q_{i}(t)$ of each constituent atom  is coupled to its own private bath, described also by a massless scalar field $\eta (y)$,  which exists in the internal spaces $y_{i}$ of each dielectric constituent atom. (This field was first introduced by Hopfield \cite{Hopfield} to provide damping in the motion of the dielectric atoms in the material.)  The inter-atomic coupling is ignored in our model, although its implementation is straightforward.

One common shared bath among the dielectric constituents, like the individual baths, also engenders dissipation in the constituents' dynamics; however, non-Markovian effects associated with the spatial nonlocality in the common bath  introduce subtleties such as a  limitation on the size of the dielectric. The dielectric can no longer be assumed to be infinite in its extension because the motion of the constituents will never reach equilibrium. For any specific constituent of the dielectric, it will always be influenced by its remotest partners at any moment if there were no restrictions on the size. Therefore for the common bath configuration, consideration of the finite medium size is absolutely necessary, and that will add in extra complications to the problem.

\subsubsection{Ordered sequence in a hierarchy of coarse-grainings}
All told, we see that the entire system consists of four interacting subsystems: the dipoles/polarizable atoms (A), scalar field (F), the dielectric medium (M) and a thermal bath (B). The presence of the latter is necessary to incorporate  dissipative behavior into the dielectric. Since of interest to us is the dynamical evolution of the dipole in the neutral atom, we treat it as our system and the other three as its environments, but importantly, in a definite ordering hierarchy. We will find the reduced dynamics of the internal degrees of freedom of the neutral atoms by tracing out the environmental variables in a physically determined order. We expect that the dielectric will change the behavior of the surrounding quantum field, and then the modified field will further alter the induced coherence between the neutral atoms. Thus we first trace out the bath variables to find a description for a dispersive, absorptive dielectric medium, and then the dielectric variables to obtain an effective description of the field in such a way that the net effects of the dielectric is captured by the dielectric function and a stochastic source. It is in this ordered manner that we reduce the formulation of four interacting subsystems into one that will give an account of the interaction between the neutral atom and the medium-modified scalar field. Subsequently we will integrate out the remaining environment variable, the modified scalar field, to acquire the full hierarchically ordered influences of the environments on the evolution of the dipole of the neutral atom, our reduced system. One can then study all the interesting features of the nonequilibrium dynamics of the reduced system.

In Paper II \cite{AFD2}, we shall define the covariant matrix of the system of $N$ atoms interacting with a medium-altered quantum field. This will enable the construction of all the physical observables of interest for the calculations of, e.g., the purity and entanglement measures between the atom and the medium. These topics will be treated in later papers of this series.

\subsection{Nested interaction of subsystems}\label{S:beei}
The full dynamics of the interacting subsystems -- dipoles of $N$ identical neutral atoms (A), ambient scalar field (F), and the dielectric medium (M) plus its accompanying thermal bath (B) -- is described by the action
\begin{equation}\label{E:djfek}
	S[\{\chi\},\phi,\varsigma,\psi]=\sum_{n=1}^{N}S_{\textsc{a}}[\chi^{(n)}]+S_{\textsc{f}}[\phi]+S_{\textsc{m}}[\varsigma]+S_{\textsc{af}}[\{\chi\},\phi]+S_{\textsc{mf}}[\varsigma,\phi]+S_{\textsc{b}}[\psi]+S_{\textsc{mb}}[\varsigma,\psi]\,,
\end{equation}
where $\{\chi\}$ is a shorthand notation for the collection of $\chi^{(1)}$, $\chi^{(2)}$, $\dots$, $\chi^{(n)}$.

The action  $S_{\textsc{a}}[\chi^{(n)}]$  of the internal degree of freedom $\chi^{(n)}$ of the $n^{\text{th}}$ neutral atom (A) describing the dynamics of the atom's dipole moment is given by,
\begin{equation}
	S_{\textsc{a}}[\chi^{(n)}]=\frac{M}{2}\int\!dt\;\Bigl(\dot{\chi}^{(n)2}-\Omega^{2}\,\chi^{(n)2}\Bigr)\,.
\end{equation}
where an overdot on a variable denotes taking its time derivative. For simplicity, we model the internal degree of freedom $\chi$ of each atom by a quantum harmonic oscillator of mass $M$ and oscillating frequency $\Omega$. The displacement $\chi$ will describe the deformation of the charge distribution by the external field inside a  neutral atom. Thus, the dipole is related to this displacement $\chi(t)$ by $\mu(t)=g_{\textsc{f}}\,\chi(t)$, where $g_{\textsc{f}}$ is the scalar charge and characterizes the coupling strength between the atom and the ambient scalar field.

The action $S_{\textsc{f}}$ of the ambient quantum field (F) $\phi$ has the form
\begin{equation}
	S_{\textsc{f}}[\phi]=\frac{1}{2}\int\!d^{4}x\; (\partial_{\mu}\phi)\,(\partial^{\mu}\phi)\,,
\end{equation}
with $x=(t,\bm{x})$. As mention earlier, the massless scalar field $\phi$ is a simplified representation of the electromagnetic field to avoid complications from its polarization and  gauge issues.

The dielectric medium (M) consists of a lattice of neutral atoms, modeled by  harmonic oscillators of the same mass $m$ and oscillating frequency $\varpi$, located at fixed positions $\bm{a}_{i}$. The action of the dielectric is given by
\begin{equation}
	S_{\textsc{m}}[\varsigma]=\frac{m}{2}\int\!dt\,d^{3}\bm{x}\sum_{i}\delta^{3}(\bm{x}-\bm{a}_{i})\,\Bigl[\dot{\varsigma}^{2}(\bm{a}_i,t)-\varpi^{2}\varsigma^{2}(\bm{a}_i,t)\Bigr]\,,
\end{equation}
where $\varsigma(\bm{a}_i,t)$ models the internal degree of freedom of the atoms of which the medium is comprised. We will assume this displacement is small compared to the distance between the medium atoms, and  much smaller than the wavelengths of the modes of our interest for the ambient field. In principle, this model can accommodate  dielectrics  containing  various species of atoms, as long as each specifies can be sufficiently well approximated by quantum harmonic oscillators of different oscillating frequencies. The latter will account for the relevant transition frequencies. In addition, the arrangement of the atoms that constitute the dielectric medium can be arbitrary, as long as they are fixed in time, so in principle various crystal structures can be included. This dielectric will be coupled to a thermal bath, modeled by another massless scalar field $\psi$ whose action is
\begin{equation}
    S_{\textsc{b}}[\psi]=\frac{1}{2}\int\!d^{4}x\; (\partial_{\mu}\psi)\,(\partial^{\mu}\psi)\,.
\end{equation}
Their interaction, described by
\begin{equation}\label{E:tikeeubfd}
	S_{\textsc{mb}}[\varsigma,\psi]=g_{\textsc{b}}\int\!dt\!\int\!d^{3}\bm{x}\;\sum_{i}\delta^{3}(\bm{x}-\bm{a}_{i})\,\varsigma(\bm{a}_i,t)\psi(\bm{x},t)\,,
\end{equation}
will give the absorptive property of the dielectric. The coupling strength $g_{\textsc{b}}$ does not need to be vanishingly weak.

Since the quantum field simultaneously couples with the atoms and the dielectric, we have two additional actions that account for the mutual interactions. The action $S_{\textsc{af}}$ describes the interaction between the field and the dipoles $\mu^{(n)}$ of the system's neutral atoms (not the ones that make up the dielectric medium), located at $\bm{z}_{n}$, 
\begin{equation}
	S_{\textsc{af}}[\{\chi\},\phi]=\sum_{n=1}^{N}\int\!dt\;\mu^{(n)}(t)\,\phi(\bm{z}_{n},t)=g_{\textsc{f}}\sum_{n=1}^{N}\int\!dt\;\chi^{(n)}(t)\,\phi(\bm{z}_{n},t)\,.
\end{equation}
while the coupling of the ambient field to the linear dielectric medium is given by
\begin{equation}\label{E:tiubfd}
	S_{\textsc{mf}}[\varsigma,\phi]=g_{\textsc{m}}\int\!dt\!\int\!d^{3}\bm{x}\;\sum_{i}\delta^{3}(\bm{x}-\bm{a}_{i})\,\varsigma(\bm{a}_i,t)\phi(\bm{x},t)\,.
\end{equation}
The coupling strength parameters $g_{\textsc{m}}$ and $g_{\textsc{f}}$ are not related.

\subsection{Nonequilibrium dynamics}
Since all the interactions are bilinear, the total action is quadratic. This Gaussian nature of the interacting systems provides two conveniences: 1) its nonequilibrium evolution can be exactly found, and 2) their quantum equations of motion are formally identical to their classical counterparts; however, the nature of the states will distinguish a quantum dynamics from a classical one, e.g., the quantum zero-point energy or the entanglement dynamics, a topic we will address in Paper II,  as an application of the current formalism.

The time evolution of the full system is described by the density matrix $\hat{\rho}$. Suppose at time $t_{i}$ we initiate the mutual interactions between subsystems, then at a later time $t$, the density matrix of the entire system is formally given by
\begin{equation}\label{E:outed}
	\hat{\rho}(t)=e^{+i\,\hat{H}(t-t_{i})}\,\hat{\rho}(t_{i})\,e^{-i\,\hat{H}(t-t_{i})}\,,
\end{equation}
where $\hat{H}$ is the Hamiltonian operator corresponding to the action $S$ of the entire system in \eqref{E:djfek}. If we subjugate the environment variables in the full density matrix by taking their traces according to the ordered hierarchy we discussed in the Introduction, we will obtain the reduced density matrix $\hat{\rho}_{A}$ for $N$ neutral atoms,
\begin{align}\label{E:dheruhd}
	\hat{\rho}_{\textsc{a}}(t)=\operatorname{Tr}_{\textsc{fmb}}\hat{\rho}(t)\,,
\end{align}
which governs the dynamics of the dipole moments of the $N$ atoms in the modified quantum field outside the dielectric body. The time evolution of the expectation value of  a quantum  quantity, say $F[\{\chi\}]$, associated with this reduced system can then be found by
\begin{equation}
	\langle F[\{\chi(t)\}]\rangle=\operatorname{Tr}_{\textsc{a}}\Bigl\{\rho_{\textsc{a}}(t)\,F[\{\chi(t)\}]\Bigr\}\,.
\end{equation}
For the Gaussian system under consideration, the reduced density matrix $\rho_{\textsc{a}}$ is easily found in the path integral formalism, as follows. Writing the right hand side of \eqref{E:outed} in terms of path integrals, we can express Eq.~\eqref{E:dheruhd} as
\begin{align}
	\rho_{\textsc{a}}(\{\chi^{(n)}_{f}\},\{\chi'^{(n)}_{f}\};t)&=\int_{\chi^{(n)}_{i}}^{\chi^{(n)}_{f}}\!\mathcal{D}\chi^{(n)}_{+}\!\int_{\chi'^{(n)}_{i}}^{\chi'^{(n)}_{f}}\!\mathcal{D}\chi^{(n)}_{-}\int\!d\phi_{f}\int_{\phi^{\vphantom{'}}_{i}}^{\phi_{f}}\!\mathcal{D}\phi_{+}\!\int_{\phi'_{i}}^{\phi_{f}}\!\mathcal{D}\phi_{-}\notag\\
    &\qquad\qquad\qquad\qquad\qquad\times\int\!d\varsigma_{f}\int_{\varsigma^{\vphantom{'}}_{i}}^{\varsigma_{f}}\!\mathcal{D}\varsigma_{+}\!\int_{\varsigma'_{i}}^{\varsigma_{f}}\!\mathcal{D}\varsigma_{-}\int\!d\psi_{f}\int_{\psi^{\vphantom{'}}_{i}}^{\psi_{f}}\!\mathcal{D}\psi_{+}\!\int_{\psi'_{i}}^{\psi_{f}}\!\mathcal{D}\psi_{-}\notag\\
	&\quad\times\exp\Bigl\{i\,S\bigl[\{\chi_{+}\},\phi_{+},\varsigma_{+},\psi_+\bigr]-i\,S\bigl[\{\chi_{-}\},\phi_{-},\varsigma_{-},\psi_-\bigr]\Bigr\}\circ\rho(t_{i})\,,\label{E:dlied}
\end{align}
where $\rho(t_{i})$ stands for the elements of the initial state of the full system and we have suppressed myriads of integrations over the initial variables by the $\circ$ operator.  The integrals of $\phi_{\pm}$, $\varsigma_{\pm}$ and $\psi_{\pm}$ in Eq~\eqref{E:dlied} are the path-integral representation of the unitary (forward/backward) time evolution operator $e^{\pm i\,H(t-t_{t})}$, corresponding to the traces over the environment variables along the time-forward and backward paths, respectively denoted by the subscript $\pm$. When a variable has the subscript $i$ or $f$, it means that the variable is evaluated at the initial time $t_i$ or the final time $t$.

One of the nice features of the Gaussian reduced system is that its asymptotic equilibrium state is independent of the choice of its initial state. If we are interested in the equilibrium configuration of the reduced system after its motion is fully relaxed, we may choose a simple initial state on the right hand sides of Eq.~\eqref{E:dlied} to evaluate the path integrals. Here, we choose the factorizable initial density matrix,
\begin{equation}\label{E:rigbfd}
	\rho(t_{i})=\rho_{\textsc{a}}(t_{i})\otimes\rho_{\textsc{f}}(t_{i})\otimes\rho_{\textsc{m}}(t_{i})\otimes\rho_{\textsc{b}}(t_{i})\,.
\end{equation}
Even though this initial state can induce undesirable artefacts like jolts \cite{HPZ92} in the reduced dynamics at very early times, they are transient effects which leave almost intractable residual influence on the final equilibrium state which is what interests us. In Eq.~\eqref{E:rigbfd}, $\rho_{\textsc{a}}(t_{i})$ and $\rho_{\textsc{m}}(t_{i})$ are composed of the Gaussian state of each dipole moment of the neutral atom, and $\rho_{\textsc{f}}(t_{i})$ is the thermal state of the massless scalar field with the temperature $\beta^{-1}$. As for the bath, we also assume that its initial density matrix $\rho_{\textsc{b}}(t_{i})$ is a thermal state with the same temperature $\beta^{-1}$. In principle, we may let the ambient field and the bath field have different initial temperatures, but that will cause extra complications due to the temperature gradient, which will introduce a thermal current between the field and the bath~\cite{HH15a}. It converts a possible equilibrium configuration into a nonequilibrium steady state configuration. Now since the initial density matrix $\rho(t_{i})$   is also Gaussian,   the path integrations in \eqref{E:dlied} can  in principle be carried out exactly, despite their rather intimidating appearance.

In summary, the task we shall undertake in this work  is to construct the appropriate influence actions for the three layers of coarse-graining, and the goal we set forth for this paper is to study the nonequilibrium stochastic dynamics of the quantum system of $N$ dipoles of the neutral atoms interacting with a quantum field which is altered by the presence of a dielectric medium. The   lowest level of coarse-graining is to see how the effect of the bath field on the dielectric, it is also the easiest since it can be described by a quantum Brownian model. The  third and highest (not including the external or mechanical dof of the atoms) level of coarse-graining is to see the effect of the dielectric-altered quantum field on our system of $N$ dipoles of the neutral atoms. This is not so difficult once we have calculated the dielectric-altered quantum field. The most involved is the middle layer, how the ambient quantum field gets modified by the absorptive dielectric and interacts with the dipoles of the atoms. Since there is already a precedent in~\cite{BH11} the task is not so daunting.  We shall indicate the apposite modifications in this work.

Since things will get complicated quickly as we proceed to deriving various effective actions, it is perhaps useful to define our nomenclature here. When the equation of motion of a subsystem is not modified by the dissipative backactions from the other subsystems, it is called the equation of motion of the `free' subsystem even though   the right hand side of the equation may contain a source term, independent of the variable of the subsystem. On the other hand, if the equation of motion of a subsystem includes the dissipative backaction, then it is called the equation of motion of the `interacting' subsystem. For any subsystem, both forms of equations of motion may show up amidst  subsequent discussions. For example, the equations of motion of the common bath $\psi$ and the dielectric atoms $\varsigma$, located at $\bm{a}_i$, are given by
\begin{align}
    \square\psi(\bm{x},t)&=g_{\textsc{b}}\sum_i\delta^{(3)}(\bm{x}-\bm{a}_i)\,\varsigma(\bm{a}_i,t)\,,\label{E:orpdh}\\
    \frac{d^2}{dt^2}\varsigma(\bm{a}_i,t)+\varpi^2\,\varsigma(\bm{a}_i,t)&=g_{\textsc{b}}\,\psi(\bm{a}_i,t)\,.\label{E:kgbfet}
\end{align}
In this context, both sets of equations are free. On the other hand, after we have solved for $\phi$ 
\begin{equation}\label{E:dbjheur}
    \psi(x)=\psi_h(x)+g_{\textsc{b}}\sum_j\int^t\!\!dt'\;G_{\mathrm{R},0}^{(\psi_h)}(\bm{a}_i,t;\bm{a}_j,t')\,\varsigma(\bm{a}_j,t')\,,
\end{equation}
and plugged this solution to Eq.~\eqref{E:kgbfet}, we obtain a new equation of motion for $\varsigma(\bm{a}_i,t)$
\begin{equation}\label{E:gbujera}
    m\,\frac{d^2}{dt^2}\varsigma(\bm{a}_i,t)+m\varpi^2\,\varsigma(\bm{a}_i,t)-g_{\textsc{b}}^2\sum_j\int^t\!\!dt'\;G_{\mathrm{R},0}^{(\psi_h)}(\bm{a}_i,t;\bm{a}_j,t')\,\varsigma(\bm{a}_j,t')=g_{\textsc{b}}\,\psi_h(\bm{a}_i,t)\,.
\end{equation}
This is the equation of motion of an interacting $\varsigma(\bm{a}_i,t)$ because the equation of motion has been modified by the interaction between $\varsigma(\bm{a}_i,t)$ and $\psi(\bm{x})$. The meaning of the nonlocal kernel $G_{\mathrm{R},0}^{(\psi_h)}$ will be explained later. The solution~\eqref{E:dbjheur} also brings up an issue that is important in the interpretations of the result, that is, the distinction between $\psi$ and $\psi_h$. The former is the complete solution to Eq.~\eqref{E:orpdh}. The latter is the homogeneous solution which satisfies the source-free equation of the free field equation, i.e., $\square \psi_h=0$. The second piece on the right hand side~\eqref{E:dbjheur} is  the inhomogeneous solution. Their characteristics will be discussed later.

\section{Stochastic Dynamics of the Medium's Constituent Atoms}\label{S:keiuet}
The stochastic dynamics of the dielectric medium's constituent atoms  can appropriately be described by the quintessential Brownian motion model of a quantum oscillator  in a common thermal bath $\psi$ ~\cite{{GS88,HPZ92,HH15b,HH18a}}. To find the influence of the shared bath on the dielectric, we carry out the integrations involving $\psi_{\pm}$, that is, we only need the part in Eq.~\eqref{E:dlied},
\begin{equation*}
    \int\!d\psi_{f}\int\!d\psi_{i}\int\!d\psi'_{i}\int_{\psi^{\vphantom{'}}_{i}}^{\psi_{f}}\!\mathcal{D}\psi_{+}\!\int_{\psi'_{i}}^{\psi_{f}}\!\mathcal{D}\psi_{-}\;\exp\Bigl\{i\,S_{\textsc{mb}}[\varsigma_+,\psi_+]+i\,S_{\textsc{b}}[\psi_+]-i\,\Bigl(+\to-\Bigr)\Bigr\}\;\rho_{\textsc{b}}(\psi_i,\psi'_i;t_{i})
\end{equation*}
where $(+\to-)$ is a shorthand notation for $S_{\textsc{mb}}[\varsigma_-,\psi_-]+S_{\textsc{b}}[\psi_-]$. {The integrations over the initial variables will be suppressed hereafter and will be implicitly denoted by the $\circ$ operator before the initial density matrix.} The above integrations give the influence functional $\mathcal{W}^{(\varsigma)}$ of the bath on the dielectric
\begin{align}\label{E:uyroiw}
    \mathcal{W}^{(\varsigma)}[\varsigma_+,\varsigma_-]&=\exp\biggl\{g_{\textsc{b}}^2\sum_{i,j}\int\!dt\,dt'\;\biggl[i\,\Delta^{(\varsigma)}(\bm{a}_i,t)\,G_{\mathrm{R},0}^{(\psi_h)}(\bm{a}_i,t;\bm{a}_j,t')\,\Sigma^{(\varsigma)}(\bm{a}_j,t')\biggr.\biggr.\notag\\
    &\qquad\qquad\qquad\qquad\qquad\qquad-\biggl.\biggl.\frac{1}{2}\,\Delta^{(\varsigma)}(\bm{a}_i,t)\,G_{\mathrm{H},0}^{(\psi_h)}(\bm{a}_i,t;\bm{a}_j,t')\,\Delta^{(\varsigma)}(\bm{a}_j,t')\biggr]\biggr\}\,.
\end{align}
Before explaining their physical meanings,  we first explain the notations. Two kernel functions $G_{\mathrm{R},0}^{(\psi_h)}(x;x')$ and $G_{\mathrm{H},0}^{(\psi_h)}(x;x')$ are respectively the retarded Green's function and the Hadamard function of the free field $\psi_h$, not to be confused with the corresponding full interacting field $\psi$.  {(See the paragraph preceding Sec.~\ref{S:keiuet}.)} They are defined by
\begin{align}\label{E:hgver}
    G_{\mathrm{R},0}^{(\psi_h)}(x;x')&=i\,\theta(t-t')\,\bigl[\hat{\psi}_h(x),\hat{\psi}_h(x')\bigr]\,,&G_{\mathrm{H},0}^{(\psi_h)}(x;x')&=\operatorname{Tr}_{\textsc{b}}\biggl(\frac{1}{2}\bigl\{\hat{\psi}_h(x),\hat{\psi}_h(x')\bigr\}\,\rho_{\textsc{b}}(t_i)\biggr)\,.
\end{align}
The two variables $\Delta_i^{(\varsigma)}(t)$ and $\Sigma_i^{(\varsigma)}(t)$ are the relative and the center of mass coordinates of the pair $\varsigma_+(\bm{a}_i,t)$, $\varsigma_-(\bm{a}_i,t)$, e.g.,
\begin{align}
    \Delta^{(\varsigma)}(\bm{a}_i,t)&=\varsigma_+(\bm{a}_i,t)-\varsigma_-(\bm{a}_i,t)\,,&\Sigma^{(\varsigma)}(\bm{a}_i,t)&=\frac{1}{2}\bigl[\varsigma_+(\bm{a}_i,t)+\varsigma_-(\bm{a}_i,t)\bigr]\,.
\end{align}
Since we need to treat multiple interacting subsystems, we will use assorted superscripts and subscripts to identify the relevant subsystems and components. For example, the superscript in $G_{\mathrm{R},0}^{(\psi_h)}(x,x')$ indicates which variable the Green's function $G_{\mathrm{R},0}^{(\psi_h)}(x,x')$ is associated with; in this case it is the  {bath field $\psi_h$}. A subscript like $\mathrm{R}$ or $\mathrm{H}$ tells the type of the Green's functions. The additional $0$ in the subscript indicates the variable in question  {satisfies the equation of motion of the free bath field, not of the interacting bath field}. We will follow these convention throughout the paper unless mentioned otherwise.

The first term in the curly brackets in Eq.~\eqref{E:uyroiw} describes three effects though they may not be obvious in this context: 1) correction or renormalization of the parameters of the dielectric, 2) dissipation in the dielectric dipoles' motion and 3) the causal influences between dipoles in the dielectric. The presence of dissipation introduces the absorptive behavior of the dielectric medium. The second term recounts the quantum fluctuations of the bath on the medium and these fluctuations will induce dipole moments of the dielectric atoms. The combined effects of the dissipation and the noise from the bath will cause the medium atoms to reach a equilibrium state under pretty general assumptions, at least for linear open systems~\cite{Ve11,HH18a}. If the coupling strength $g_{\textsc{b}}$ is very weak, the end state will be a thermal state having the same temperature as the bath. Otherwise, if the coupling constant has finite strength, then the final state will not possess a Gibbs form; it will depend on $g_{\textsc{b}}$ and other parameters of the dielectric atom in general~\cite{HH18a,HH19a,HH19b}.

\subsection{Effective actions of the dielectric medium}
The expression Eq.~\eqref{E:uyroiw} will be combined with the action $S_{\textsc{m}}$ and $S_{\textsc{mf}}$ to form the coarse-grained effective action $S_{\textsc{cgea}}^{(\varsigma)}$ \cite{cgea,cgea1} of the dielectric medium
\begin{equation}\label{E:jeoiu}
    S_{\textsc{cgea}}^{(\varsigma)}[\varsigma_+,\varsigma_-]=S_{\textsc{m}}[\varsigma_+]+S_{\textsc{mf}}[\varsigma_+,\phi_+]-S_{\textsc{m}}[\varsigma_-]+S_{\textsc{mf}}[\varsigma_-,\phi_-]+S^{(\varsigma)}_{\textsc{if}}[\varsigma_{+},\varsigma_{-}]
\end{equation}
with $S^{(\varsigma)}_{\textsc{if}}[\varsigma_{+},\varsigma_{-}]=-i\,\ln\mathcal{W}^{(\varsigma)}[\varsigma_{+},\varsigma_{-}]$. The action \eqref{E:jeoiu} will be used to find the influence action of the medium on the ambient field $\phi$ in the next section.

{As can be seen from Eq.~\eqref{E:uyroiw},  the coarse-grained effective action $S_{\textsc{cgea}}^{(\varsigma)}$ is in general a complex action, so it often causes difficulty in interpretation of its implications. We may use the the Feynman-Vernon scheme to transform it into a real action.} Observe that the real part of the influence functional $\mathcal{W}^{(\varsigma)}$ in Eq.~\eqref{E:uyroiw} as
\begin{align}\label{E:vdk}
    &\quad\exp\biggl\{-\frac{g_{\textsc{b}}^2}{2}\sum_{i,j}\int\!dt'\,dt''\;\Delta^{(\varsigma)}(\bm{a}_i,t')\,G_{\mathrm{H},0}^{(\psi_h)}(\bm{a}_i,t';\bm{a}_j,t'')\,\Delta^{(\varsigma)}(\bm{a}_j,t'')\biggr\}\notag\\
    &=\int\!\mathcal{D}\xi_0^{(\psi_h)}\;\mathbb{P}_{\psi_h}[\xi_0^{(\psi_h)}]\,\exp\biggl[i\,g_{\textsc{b}}\sum_i\int\!dt'\;\Delta^{(\varsigma)}(\bm{a}_i,t')\,\xi_0^{(\psi_h)}(\bm{a}_i,t')\biggr]\,,
\end{align}
by introducing a stochastic averaging over the noise variable $\xi_0^{(\psi_h)}$. The choice of the probability distribution functional $\mathbb{P}_{\psi}[\xi_0^{(\psi_h)}]$ is such that
\begin{align}
    \int\!\mathcal{D}\xi_0^{(\psi_h)}\;\mathbb{P}_{\psi_h}[\xi_0^{(\psi_h)}]\,\xi_0^{(\psi_h)}(\bm{x},t)&=0\,,\\
    \int\!\mathcal{D}\xi_0^{(\psi_h)}\;\mathbb{P}_{\psi_h}[\xi_0^{(\psi)}]\,\xi_0^{(\psi_h)}(\bm{x},t)\,\xi_0^{(\psi_h)}(\bm{x}',t')&=G_{\mathrm{H},0}^{(\psi_h)}(\bm{x},t;\bm{x}',t')\,,\label{E:bwiys}
\end{align}
and no more. The higher even moments will be expanded in terms of the second moments, following the Wick expansion. Thus the probability distribution functional $\mathbb{P}_{\psi_h}[\xi_0^{(\psi_h)}]$ is a Gaussian. It will be seen that the stochastic noise $\xi_0^{(\psi_h)}(x)$ in a sense corresponds to the bath field $\psi_h(x)$, satisfying a source-free, free wave equation, and thus will account for the quantum noise from the free bath field.  {In fact, we can interpret an expression like $\displaystyle\int\mathcal{D}\xi_0^{(\psi_h)}\;\mathbb{P}_{\psi_h}[\xi_0^{(\psi_h)}]\,(\cdots)$ as the stochastic or statistical average $\langle\cdots\rangle_{\xi_0^{(\psi_h)}}$ over classical noise $\xi_0^{(\psi_h)}$ with the probability distribution functional $\mathbb{P}_{\psi_h}[\xi_0^{(\psi_h)}]$. This will come  in handy for computing the physical observables hereafter.}

This allows us to write the coarse-grained effective action of the dielectric medium $S_{\textsc{cgea}}^{(\varsigma)}[\varsigma_+,\varsigma_-]$ in terms of the stochastic effective action \cite{JH1,HH15a} of the dielectric medium $S_{\textsc{se}}^{(\varsigma)}[\varsigma_+,\varsigma_-;\xi_0^{(\psi_h)}]$ by
\begin{align}
    e^{i\,S_{\textsc{cgea}}^{(\varsigma)}[\varsigma_+,\varsigma_-]}=\int\!\mathcal{D}\xi_0^{(\psi_h)}\;\mathbb{P}_{\psi_h}[\xi_0^{(\psi_h)}]\,e^{S_{\textsc{se}}^{(\varsigma)}[\varsigma_+,\varsigma_-;\xi_0^{(\psi_h)}]}\,,
\end{align}
with, by construction,
\begin{align}\label{E:bfiers}
    S_{\textsc{se}}^{(\varsigma)}[\varsigma_+,\varsigma_-;\xi_0^{(\psi_h)}]&=S_{\textsc{m}}\bigl[\varsigma_{+}\bigr]+S_{\textsc{mf}}[\varsigma_{+},\phi_{+}]-S_{\textsc{m}}\bigl[\varsigma_{-}\bigr]-S_{\textsc{mf}}[\varsigma_{-},\phi_{-}]\notag\\
    &\qquad\qquad+g_{\textsc{b}}^2\sum_{i,j}\int\!dt\,dt'\;\Delta^{(\varsigma)}(\bm{a}_i,t)\,G_{\mathrm{R},0}^{(\psi_h)}(\bm{a}_i,t;\bm{a}_j,t')\,\Sigma^{(\varsigma)}(\bm{a}_j,t')\notag\\
    &\qquad\qquad\qquad\qquad+g_{\textsc{b}}\sum_i\int\!dt'\;\Delta^{(\varsigma)}(\bm{a}_i,t')\,\xi_0^{(\psi_h)}(\bm{a}_i,t')\,.
\end{align}
Clearly, this action is real like a typical action. Thus in principle we can take the functional variation of $S_{\textsc{se}}^{(\varsigma)}[\varsigma_+,\varsigma_-;\xi_0^{(\psi_h)}]$ to find an equation of motion. Doing so with respect to $\Delta^{(\varsigma)}$ yields the Langevin equation,
\begin{equation}\label{E:sltie}
    \ddot{\varsigma}(\bm{a}_i,t)+\varpi^2\varsigma(\bm{a}_i,t)-\frac{g_{\textsc{b}}^2}{m}\sum_j\int_0^t\!dt'\;G_{\mathrm{R},0}^{(\psi_h)}(\bm{a}_i,t;\bm{a}_j,t')\,\varsigma(\bm{a}_j,t')=\frac{g_{\textsc{b}}}{m}\,\xi_0^{(\psi_h)}(\bm{a}_i,t)+\frac{g_{\textsc{m}}}{m}\,\phi(\bm{a}_i,t)\,,
\end{equation}
after setting $\Delta^{(\varsigma)}=0$.   {Note here $\phi$ is a fully interacting field, unlike $\xi_0^{(\psi_h)}$, a doppelg\"anger of the free bath field $\psi_h$.} In treating interacting  multipartite systems, it is very important to distinguish a free variable from an interacting variable in order to avoid confusion and misinterpretation of the physical implications. {Note here 1) a correction to Eq. (24) of Ref.~\cite{BH11}, where a Lorentz-like force $g_{\textsc{m}}\,\phi(\bm{a}_i,t)$ at the location of the dielectric atom is missing, and 2) the difference: there are no causal influences between the dielectric atoms in ~\cite{BH11} where because each dielectric atom has its private bath whereas a common bath for all dielectric atoms is assumed here.}

From this Langevin equation we have sought after, the physical meanings of each term in the coarse-grained effective actions $S_{\textsc{cgea}}^{(\varsigma)}$ outlined above can now be clearly seen.  Eq.~\eqref{E:sltie} takes the form of the equation of motion for a driven damped oscillator. The nonlocal term for the case $j=i$ will introduce a correction to the oscillating frequency from $\varpi$ to $\varpi_{\textsc{r}}$, and a radiation damping,  well explained in, e.g.,  Ch.14 of~\cite{Ja98} (see also \cite{Ro90,Ya06,Sp04}), which gives the reactive force to dipole's motion. When $j\neq i$, due to the presence of the retarded Green's function of the free field $\psi_h$, it accounts for the causal influence of the dielectric atom at $\bm{a}_j$ at time $t'$ on another dielectric atom at $\bm{a}_i$ at time $t$, mediated by the bath. Thus it is a spatial nonMarkovian effect.  {The role of the bath is better seen from the solution to Eq.~\eqref{E:sltie}
\begin{equation}\label{E:geres}
    \varsigma(\bm{a}_i,t)=\varsigma_h(\bm{a}_i,t)+\sum_j\int_0^t\!dt'\;G_{\mathrm{R}}^{(\varsigma_h)}(\bm{a}_i,t;\bm{a}_j,t')\Bigl[\frac{g_{\textsc{b}}}{m}\,\xi_0^{(\psi_h)}(\bm{a}_j,t')+\frac{g_{\textsc{m}}}{m}\,\phi(\bm{a}_j,t')\Bigr]
\end{equation}
where $\varsigma_h$ depends only on the initial conditions, but the second term, the inhomogeneous solution, relies on the value of $\xi_0^{(\psi_h)}$ at the site $\bm{a}_j$ at earlier times $t'$}, {due to the retarded Green's function $G_{\mathrm{R}}^{(\varsigma_h)}(\bm{a}_i,t;\bm{x}',t')$, which measures the causal response of $\varsigma(\bm{a}_i,t)$ from $\bm{x}'$ at time $t'$.} This mutual influence introduces memory to the dynamics of the dielectric dipoles, adding complicated behavior to their time evolution~\cite{HH15b,HH18a,HH19a}.

We then can derive the influence functional \eqref{E:reuthsiud} of the dielectric on the ambient field, starting from {the stochastic effective action $S_{\textsc{se}}^{(\varsigma)}[\varsigma_+,\varsigma_-;\xi_0^{(\psi_h)}]$ or the coarse-grained effective action $S_{\textsc{cgea}}^{(\varsigma)}[\varsigma_+,\varsigma_-]$ of the dielectric. They will give a slightly different but equivalent influence action of the dielectric.}

\subsection{Influence action of the interlocutor ambient quantum field, a preview}\label{S:aket}
A dielectric medium comprises a huge number of quantum degrees of freedom, but their overall influence can be succinctly summarized into a dielectric function with accompanying stochastic noise current. By eliminating the explicit appearance of the medium's degrees of freedom in the total density matrix $\rho$, we will obtain the coarse-grained effective action $S_{\textsc{cgea}}^{(\phi)}$ for the field~\cite{cgea}. In general such a coarse-grained effective action may contain an infinite number of terms and could be non-local in space and time. However, in our assumption of Gaussian form, because the dielectric medium consists of a lattice of non-overlapping, non-interacting harmonic oscillators, all of which are linearly coupled to the field, their contributions to the coarse-grained effective action are local in space and quadratic.

This leads to the generating functional $\mathcal{W}^{(\phi)}[\phi_{+},\phi_{-}]$ measuring the influences of the dielectric on the field, which can be expressed conveniently in terms of an influence action~\cite{FeyVer},
\begin{align}
	\mathcal{W}^{(\phi)}[\phi_{+},\phi_{-}]&=\int\!\mathcal{D}\xi_0^{(\psi_h)}\;\mathbb{P}_{\psi_h}[\xi_0^{(\psi_h)}]\int\!d\varsigma_{f}\int_{\varsigma^{\vphantom{'}}_{i}}^{\varsigma_{f}}\!\mathcal{D}\varsigma_{+}\!\int_{\varsigma'_{i}}^{\varsigma_{f}}\!\mathcal{D}\varsigma_{-}\;\exp\Bigl\{i\,S_{\textsc{se}}^{(\varsigma)}[\varsigma_+,\varsigma_-;\xi_0^{(\psi_h)}]\Bigr\}\circ\rho_{\textsc{m}}(t_{i})\label{E:vbxir}\\
 &=\int\!d\varsigma_{f}\int_{\varsigma^{\vphantom{'}}_{i}}^{\varsigma_{f}}\!\mathcal{D}\varsigma_{+}\!\int_{\varsigma'_{i}}^{\varsigma_{f}}\!\mathcal{D}\varsigma_{-}\;\exp\Bigl\{i\,S_{\textsc{cgea}}^{(\varsigma)}[\varsigma_+,\varsigma_-]\Bigr\}\circ\rho_{\textsc{m}}(t_{i})\,.\label{E:heryes}
\end{align}
 {The approach based on Eq.~\eqref{E:heryes} has been shown in~\cite{CRV}, so here we will take Eq.~\eqref{E:vbxir} as the starting point. Then the derived influence action $S^{(\phi)}_{\textsc{if}}$ on the ambient field $\phi$ will contain the stochastic noise. To avoid confusion, we will called it `stochastic' influence action and denote it by $S^{(\phi)}_{\textsc{sif}}$.}

{
To carry out the functional integration over $\varsigma_{\pm}$, observe that at this lowest level in the hierarchy the stochastic noise $\xi_0^{(\psi_h)}$ can be treated as external, so we can combine the interaction terms $S_{\textsc{mf}}[\varsigma_{+},\phi_{+}]-S_{\textsc{mf}}[\varsigma_{-},\phi_{-}]$ in Eq.~\eqref{E:bfiers} and the contribution related to $\xi_0^{(\psi_h)}$ into
\begin{align}
    g_{\textsc{m}}\sum_i\int^t\!\!dt'\Bigl[\Sigma^{(\varsigma)}(\bm{a}_i,t)\,\Delta^{(\mathscr{J})}(\bm{a}_i,t)+\Delta^{(\varsigma)}(\bm{a}_i,t)\,\Sigma^{(\mathscr{J})}(\bm{a}_i,t)\Bigr]
\end{align}
where $g_{\textsc{m}}\,\Sigma^{(\mathscr{J})}(x)=g_{\textsc{m}}\,\Sigma^{(\phi)}(x)+g_{\textsc{b}}\,\xi_0^{(\psi_h)}(x)$,  $g_{\textsc{m}}\,\Delta^{(\mathscr{J})}(x)=g_{\textsc{m}}\,\Delta^{(\phi)}(x)$ with $g_{\textsc{m}}\,\mathscr{J}_{\pm}(x)=g_{\textsc{m}}\,\phi_{\pm}+g_{\textsc{b}}\,\xi_0^{(\psi_h)}(x)$. In this way, we can put $S_{\textsc{se}}^{(\varsigma)}[\varsigma_+,\varsigma_-;\xi_0^{(\psi_h)}]$ in Eq.~\eqref{E:vbxir} in the form
\begin{align}\label{E:ljbieur}
    S_{\textsc{se}}^{(\varsigma)}[\varsigma_+,\varsigma_-;\xi_0^{(\psi_h)}]=\int^{t}\!\!dt'\;\biggl\{-\frac{1}{2}\,\bm{Z}^{T}(t')\cdot\bm{\mathfrak{L}}'\cdot\bm{Z}(t)+g_{\textsc{m}}\,\bm{Z}^{T}(t)\cdot\bm{K}\cdot\bm{\mathscr{J}}(t)\biggr\}\,,
\end{align}
where $\bm{Z}^{T}=(\Sigma^{(\varsigma)},\,\Delta^{(\varsigma)})$, $\mathscr{L}=m^2\dfrac{d^2}{dt^2}+m^2\varpi^2$,
\begin{align}
	\bm{\mathscr{L}}&=\begin{pmatrix}0&\mathscr{L}\\\mathscr{L}&0\end{pmatrix}\,,&\bm{K}&=\begin{pmatrix}0&1\\1&0\end{pmatrix}\,,&\bm{\mathscr{J}}&=\begin{pmatrix}\Sigma^{(\mathscr{J})}\\\Delta^{(\mathscr{J})}\end{pmatrix}\,,
\end{align}
and
\begin{align*}
	\bm{\mathfrak{L}}'\cdot\bm{Z}(\bm{a}_i,t)=\bm{\mathfrak{L}}\cdot\bm{Z}(\bm{a}_i,t)-g_{\textsc{b}}^{2}\sum_j\int^{t}\!\!dt'\begin{pmatrix}0 &G^{(\phi_{h})}_{\mathrm{A}}(\bm{a}_i,t;\bm{a}_j,t')\\G^{(\phi_{h})}_{\mathrm{R}}(\bm{a}_i,t;\bm{a}_j,t') &0\end{pmatrix}\cdot\bm{Z}(\bm{a}_j,t')\,.
\end{align*}
Following the standard treatment of the Schwinger-Keldysh, closed-time-path or in-in formalism~\cite{ctp,ctp1,ctp2,CalHu08}, we find that the stochastic influence functional is given by
\begin{align}
    e^{i\,S^{(\phi)}_{\textsc{sif}}[\phi_{+},\phi_{-};\xi_0^{(\psi_h)}]}&=\int\!d\varsigma_{f}\int_{\varsigma^{\vphantom{'}}_{i}}^{\varsigma_{f}}\!\mathcal{D}\varsigma_{+}\!\int_{\varsigma'_{i}}^{\varsigma_{f}}\!\mathcal{D}\varsigma_{-}\;\exp\Bigl\{i\,S_{\textsc{se}}^{(\varsigma)}[\varsigma_+,\varsigma_-;\xi_0^{(\psi_h)}]\Bigr\}\circ\rho_{\textsc{m}}(t_{i})\\
    &=\exp\biggl\{i\,\frac{g_{\textsc{m}}^{2}}{2}\sum_{i,j}\int^{t}\!\!dt'\!\int^{t}\!\!dt''\;\bm{\mathscr{J}}^{T}(\bm{a}_i,t')\cdot\bm{K}\cdot\bm{\mathscr{L}}'^{-1}(\bm{a}_i,t';\bm{a}_j,t'')\cdot\bm{K}\cdot\bm{\mathscr{J}}(\bm{a}_j,t'')\biggr\}\,,\notag
\end{align}
where
\begin{equation}
    \bm{\mathscr{L}}'^{-1}(x,x')=\begin{pmatrix}i\,G^{(\chi_{h})}_{\mathrm{H}}(x,x') &G^{(\chi_{h})}_{\mathrm{R}}(x,x')\\[6pt]G^{(\chi_{h})}_{\mathrm{A}}(x,x') &0\end{pmatrix}\,.
\end{equation}
Explicitly, we have}
\begin{align}\label{E:reuthsiud}	
	S^{(\phi)}_{\textsc{sif}}[\phi_{+},\phi_{-};\xi_0^{(\psi_h)}]&=g_{\textsc{m}}^{2}\sum_{i,j}\int^t\!\!dt'\,dt''\;\Bigl[\Delta^{(\phi)}(\bm{a}_{i},t')\,G_{\mathrm{R}}^{(\varsigma_h)}(\bm{a}_i,t';\bm{a}_j,t'')\,\Sigma^{(\phi)}(\bm{a}_{j},t'')\Bigr.\notag\\
 &\qquad\qquad\qquad\qquad\qquad\qquad\qquad+\Bigl.\frac{i}{2}\,\Delta^{(\phi)}(\bm{a}_{i},t')\,G^{(\varsigma_h)}_{\mathrm{H}}(\bm{a}_i,t';\bm{a}_j,t'')\,\Delta^{(\phi)}(\bm{a}_{j},t'')\Bigr]\notag\\
    &\qquad\qquad+g_{\textsc{m}}g_{\textsc{b}}\sum_{i,j}\int^t\!\!dt'\,dt''\;\Delta^{(\phi)}(\bm{a}_{i},t')\,G_{\mathrm{R}}^{(\varsigma_h)}(\bm{a}_i,t';\bm{a}_j,t'')\,\xi_0^{(\psi_h)}(\bm{a}_j,t'')\,,
\end{align}
where $\Sigma^{(\phi)}=(\phi_{+}+\phi_{-})/2$ and $\Delta^{(\phi)}=\phi_{+}-\phi_{-}$. Here the medium's retarded and Hadamard functions  {$G_{\mathrm{R}}^{(\varsigma_h)}$ and $G_{\mathrm{H}}^{(\varsigma_h)}$ respectively are defined in relation to the homogeneous part $\varsigma_h$ of} the interacting $\varsigma$, including the influences of the bath, in a way similar to those in Eq.~\eqref{E:hgver}.  {The contribution of the inhomogeneous contribution from the ``forces" on the right hand side of Eq.~\eqref{E:sltie} does not enter the definitions of $G_{\mathrm{R}}^{(\varsigma_h)}$ and $G_{\mathrm{H}}^{(\varsigma_h)}$.

\subsection{Dissipation and noise kernels of the medium atoms}
The exact forms of $G^{(\varsigma_h)}_{\mathrm{R}}(\bm{a}_i,t;\bm{a}_j,t')$ and $G^{(\varsigma_h)}_{\mathrm{H}}(\bm{a}_i,t;\bm{a}_j,t')$ are in general rather complicated due to the mutual causal influences between the dielectric atoms. However, in the case where those influences can be neglected, $G^{(\varsigma_h)}_{\mathrm{R}}(\bm{a}_i,t;\bm{a}_j,t')$ takes a simple form
\begin{align}\label{E:aesobde}
	G^{(\varsigma_h)}_{\mathrm{R}}(t,t')&=\frac{1}{m\varpi_{b}}\,\theta(t-t')\,e^{-\gamma_{\textsc{b}}(t-t')}\,\sin\varpi_{b}(t-t')\,,
\end{align}
where $\gamma_{\textsc{b}}=g_{\textsc{b}}^2/(8\pi m)$ is the damping constant, gauging the strength of the absorption induced by the medium-bath interaction, and $\varpi_{b}=\sqrt{\varpi_{\textsc{b}}^2-\gamma_{\textsc{b}}^2}$, in which $\varpi_{\textsc{b}}$ is the physical frequency associated with the dipole of the dielectric atom, corrected by the dielectric-bath interaction. In this case the form of $G^{(\varsigma_h)}_{\mathrm{R}}(t,t')$ is independent of the locations of dielectric atoms $\bm{a}_i$ because we also have assumed identical dielectric atoms so far  {and have neglected the mutual influences. The results obtained in the previous section} can be easily generalized to the cases that the dielectric comprises assorted species of simple atoms.

Let us take a closer look at \eqref{E:reuthsiud}. The first two terms there would make up what we would expect for an influence action, but we have an additional term depending on the stochastic noise $\xi^{(\psi_h)}_0$ from the lower tier, i.e., the thermal bath. This is what begins to be interesting. To get back to the bona fide form of the influence action for the field $S^{(\phi)}_{\textsc{if}}[\phi_{+},\phi_{-}]=-i\,\ln\mathcal{W}^{(\phi)}[\phi_{+},\phi_{-}]$ in  \eqref{E:vbxir}, we would continue to evaluate the statistical ensemble average over the noise $\xi_0^{(\psi_h)}$ in \eqref{E:vbxir}
\begin{align}\label{E:dnfweri}
    &\hphantom{=}\mathcal{W}^{(\phi)}[\phi_{+},\phi_{-}]\notag\\
    &=\int\!\mathcal{D}\xi_0^{(\psi_h)}\;\mathbb{P}_{\psi_h}[\xi_0^{(\psi_h)}]\,e^{i\,S^{(\phi)}_{\textsc{sif}}[\phi_{+},\phi_{-};\xi_0^{(\psi_h)}]}\notag\\
    &=\exp\biggl\{i\,g_{\textsc{m}}^{2}\sum_{i,j}\int^t\!\!dt'\,dt''\;\Bigl[\Delta^{(\phi)}(\bm{a}_{i},t')\,G_{\mathrm{R}}^{(\varsigma_h)}(\bm{a}_i,t';\bm{a}_j,t'')\,\Sigma^{(\phi)}(\bm{a}_{j},t'')\Bigr.\biggr.\notag\\
 &\qquad\qquad\qquad\qquad\qquad\qquad\qquad\qquad\qquad\qquad\quad+\biggl.\Bigl.\frac{i}{2}\,\Delta^{(\phi)}(\bm{a}_{i},t')\,G^{(\varsigma_h)}_{\mathrm{H}}(\bm{a}_i,t';\bm{a}_j,t'')\,\Delta^{(\phi)}(\bm{a}_{j},t'')\Bigr]\biggr\}\notag\\
	&\times\exp\biggl\{-\frac{g_{\textsc{m}}g_{\textsc{b}}^{2}}{2}\sum_{i,j}\sum_{k,l}\int^{t}\!\!dt'\!\int^{t}\!\!dt''\!\int^{t'}\!\!d\tau'\!\int^{t''}\!\!d\tau''\;\Delta^{(\phi)}(\bm{a}_i,t')\,G_{\mathrm{R}}^{(\varsigma_{h})}(\bm{a}_i,t';\bm{a}_k,\tau')\,G_{\mathrm{H},0}^{(\psi_{h})}(\bm{a}_k,\tau';\bm{a}_l,\tau'')\biggr.\notag\\
 &\qquad\qquad\qquad\qquad\qquad\qquad\qquad\qquad\qquad\qquad\qquad\times\biggl.G_{\mathrm{R}}^{(\varsigma_{h})}(\bm{a}_j,t'';\bm{a}_l,\tau'')\,\Delta^{(\phi)}(\bm{a}_j,t'')\biggr\}\,,
\end{align}
where we have used the Feynman-Vernon (FV) identity Eq.~\eqref{E:vdk} in the reverse order. {Recall what we emphasized from the beginning  a very important characteristic of the FV formalism for Gaussian systems, namely, the noise obtained via the FV identity, suffers no loss of information in the quantum realm in transition to a stochastic representation. Thus the reverse transform reinstated all quantum features untarnished. This is not true for many popular quantum-stochastic schemes.}

Define
\begin{align}\label{E:vbkeru}
	G_{\mathrm{H}}^{(\varsigma)}(\bm{a}_i,t';,\bm{a}_j,t'')&=G_{\mathrm{H}}^{(\varsigma_{h})}(\bm{a}_i,t';,\bm{a}_j,t'')\\
 &+g_{\textsc{b}}^{2}\sum_{k,l}\int^{t'}\!\!d\tau'\!\int^{t''}\!\!d\tau''\;G_{\mathrm{R}}^{(\varsigma_{h})}(\bm{a}_i,t';\bm{a}_k,\tau')\,G_{\mathrm{R}}^{(\varsigma_{h})}(\bm{a}_j,t'';\bm{a}_l,\tau'')\,G_{\mathrm{H},0}^{(\phi_{h})}(\bm{a}_k,\tau';\bm{a}_l,\tau'')\,,\notag
\end{align}
and we claim that this is the Hadamard function of the complete, interacting $\varsigma$, in contrast to $\,G_{\mathrm{H}}^{(\varsigma_{h})}(\bm{a}_i,t';,\bm{a}_j,t'')$, which is the Hadamard function of $\varsigma_{h}$. In this way, we arrive at the influence action of the field $S^{(\phi)}_{\textsc{if}}[\phi_{+},\phi_{-}]$,
\begin{align}\label{E:wojrt}
    S^{(\phi)}_{\textsc{if}}[\phi_{+},\phi_{-}]&=g_{\textsc{m}}^{2}\sum_{i,j}\int^t\!\!dt'\,dt''\;\Bigl[\Delta^{(\phi)}(\bm{a}_{i},t')\,G_{\mathrm{R}}^{(\varsigma_h)}(\bm{a}_i,t';\bm{a}_j,t'')\,\Sigma^{(\phi)}(\bm{a}_{j},t'')\Bigr.\notag\\
 &\qquad\qquad\qquad\qquad\qquad\qquad\quad+\Bigl.\frac{i}{2}\,\Delta^{(\phi)}(\bm{a}_{i},t')\,G^{(\varsigma)}_{\mathrm{H}}(\bm{a}_i,t';\bm{a}_j,t'')\,\Delta^{(\phi)}(\bm{a}_{j},t'')\Bigr]\,.
\end{align}
Note the subtle differences (marked in the superscripts) of the variables used for the dissipation kernel and the noise kernel.
}

{
To certify our claim in Eq.~\eqref{E:vbkeru}, we start from the quantum equation of motion  the operator $\hat{\varsigma}$,
\begin{equation}	
	m\,\frac{d^2}{dt^2}\hat{\varsigma}(\bm{a}_i,t)+m\varpi^2\hat{\varsigma}(\bm{a}_i,t)-g_{\textsc{b}}^{2}\int^{t}\!\!dt'\;G_{\mathrm{R},0}^{(\psi_{h})}(\bm{a}_i,t;\bm{a}_j,t')\,\hat{\varsigma}(\bm{a}_j,t')=g_{\textsc{b}}\,\hat{\psi}_h(\bm{a}_i,t)\,,
\end{equation}
which is the quantum counterpart of Eq.~\eqref{E:gbujera}. Its full solution is given by
\begin{equation}\label{E:qwsdf}
    \hat{\varsigma}(\bm{a}_i,t)=\hat{\varsigma}_h(\bm{a}_i,t)+g_{\textsc{b}}\sum_k\int^t\!\!d\tau\;G_{\mathrm{R}}^{(\varsigma_h)}(\bm{a}_i,t;\bm{a}_k,\tau)\,\hat{\psi}_h(\bm{a}_k,\tau)\,.
\end{equation}
Then by a similar construction to Eq.~\eqref{E:hgver}, we obtain
\begin{align}\label{E:bverus}
    G_{\mathrm{H}}^{(\varsigma)}(\bm{a}_i,t;\bm{a}_j,t')&=\operatorname{Tr}_{\textsc{m}\textsc{b}}\biggl(\frac{1}{2}\bigl\{\hat{\varsigma}(\bm{a}_i,t),\hat{\varsigma}(\bm{a}_j,t')\bigr\}\,\rho_{\textsc{m}}(t_i)\otimes\rho_{\textsc{b}}(t_i)\biggr)\notag\\
    &=G_{\mathrm{H}}^{(\varsigma_h)}(\bm{a}_i,t;\bm{a}_j,t')+g_{\textsc{b}}^2\sum_{k,l}\int^t\!\!d\tau\!\int^{t'}\!\!d\tau'\;G_{\mathrm{R}}^{(\varsigma_h)}(\bm{a}_i,t;\bm{a}_k,\tau)\,G_{\mathrm{R}}^{(\varsigma_h)}(\bm{a}_j,t';\bm{a}_l,\tau')\notag\\
    &\qquad\qquad\qquad\qquad\qquad\qquad\qquad\qquad\times G_{\mathrm{H},0}^{(\psi_h)}(\bm{a}_k,\tau;\bm{a}_l,\tau')\,.
\end{align}
This standard quantum-mechanical construction gives exactly Eq.~\eqref{E:vbkeru}.}

Observe that the expression of the influence action $S^{(\phi)}_{\textsc{if}}[\phi_{+},\phi_{-}]$ in Eq.~\eqref{E:wojrt} prompts the application of the Feynman-Vernon identity. Introduce a probability distribution functional $\mathbb{P}_{\varsigma}[\xi^{(\varsigma)}]$ such that
\begin{align}\label{E:qbkder}
	&\quad\exp\biggl\{-\frac{g_{\textsc{m}}^{2}}{2}\!\sum_{i,j}\int^t\!\!dt'\,dt''\;\Delta^{(\phi)}(\bm{a}_{i},t')\,G^{(\varsigma)}_{\mathrm{H}}(\bm{a}_i,t';\bm{a}_j,t'')\,\Delta^{(\phi)}(\bm{a}_{j},t'')\Bigr]\biggr\}\notag\\
	&=\int\!\mathcal{D}\xi^{(\varsigma)}\;\mathbb{P}_{\varsigma}[\xi^{(\varsigma)}]\,\exp\biggl\{i\,g_{\textsc{m}}\sum_i\int^t\!\!dt'\;\Delta^{(\phi)}(\bm{a}_{i},t')\,\xi^{(\varsigma)}(\bm{a}_{i},t')\biggr\}\,,
\end{align}
with the conditions on the moments
\begin{align}
	\int\!\mathcal{D}\xi^{(\varsigma)}\;\mathbb{P}_{\varsigma}[\xi^{(\varsigma)}]\,\xi^{(\varsigma)}(x)&=0\,,&\int\!\mathcal{D}\xi^{(\varsigma)}\;\mathbb{P}_{\varsigma}[\xi^{(\varsigma)}]\,\xi^{(\varsigma)}(x)\,\xi^{(\varsigma)}(x')&=G_{\mathrm{H}}^{(\varsigma)}(x;x')\,.
\end{align}
Any higher odd moments will vanish but the even moments can be obtained by the Wick expansion. Then we can write the influence functional Eq.~\eqref{E:dnfweri} as
\begin{align}\label{E:ifdghn}
	\mathcal{W}^{(\phi)}[\phi_{+},\phi_{-}]&=\exp\biggl\{i\,S_{\textsc{if}}^{(\phi)}[\phi_{+},\phi_{-}]\biggr\}\notag\\
	&=\int\!\mathcal{D}\xi^{(\varsigma)}\;\mathbb{P}_{\varsigma}[\xi^{(\varsigma)}]\,\exp\biggl\{i\,e^{2}\!\int^{t}\!\!dt'\int^{t}\!\!dt''\;\Delta^{(\phi)}(\bm{a}_i,t')\,G_{\mathrm{R}}^{(\varsigma_{h})}(\bm{a}_i,t';\bm{a}_j,t'')\,\Sigma^{(\phi)}(\bm{a}_j,t'')\biggr.\notag\\
    &\qquad\qquad\qquad\qquad\qquad\qquad+\biggl.i\,g_{\textsc{m}}\!\int^{t}\!\!dt'\;\Delta^{(\phi)}(\bm{a}_i,t')\,\xi^{(\varsigma)}(\bm{a}_i,t')\biggr\}\,.
\end{align}
Consequently, we make the identification $\xi^{(\varsigma)}$ as 
\begin{align}\label{E:wewsdw}
	\xi^{(\varsigma)}(x)&=\xi^{(\varsigma_{h})}(x)+g_{\textsc{b}}\int^{x}\!\!d^{4}y\;G_{\mathrm{R}}^{(\varsigma_{h})}(x,y)\,\xi_{0}^{(\psi_h)}(y)\,.
\end{align}

To see the significance of this result, let us shift our attention back to the stochastic influence action $S^{(\phi)}_{\textsc{sif}}[\phi_{+},\phi_{-};\xi_0^{(\psi_h)}]$ of the medium on the ambient field in Eq.~\eqref{E:reuthsiud}. We can alternatively apply the Feynman-Vernon identity to $G^{(\varsigma_h)}_{\mathrm{H}}(\bm{a}_i,t';\bm{a}_j,t'')$ on the right hand side by introducing  another probability distribution functional $\mathbb{P}_{\varsigma_{h}}[\xi^{(\varsigma_{h})}]$
\begin{align*}
	&\quad\exp\biggl\{-\frac{g_{\textsc{m}}^{2}}{2}\sum_{i,j}\int^{t}\!\!dt'\!\int^{t}\!\!dt''\;\Delta^{(\phi)}(\bm{a}_i,t')\,G^{(\varsigma_h)}_{\mathrm{H}}(\bm{a}_i,t';\bm{a}_j,t'')\,\Delta^{(\phi)}(\bm{a}_j,t'')\biggr\}\notag\\
	&=\int\!\mathcal{D}\xi^{(\varsigma_{h})}\;\mathbb{P}_{\varsigma_{h}}[\xi^{(\varsigma_{h})}]\,\exp\biggl\{i\,g_{\textsc{m}}\sum_i\int^{t}\!\!dt'\;\Delta^{(\phi)}(\bm{a}_i,t')\,\xi^{(\varsigma_{h})}(\bm{a}_i,t')\biggr\}\,,
\end{align*}
such that the influence functional $\mathcal{W}^{(\phi)}[\phi_{+},\phi_{-}]$ in Eq.~\eqref{E:dnfweri} can be alternatively cast in the form
\begin{align}\label{E:beiurw}
    \mathcal{W}^{(\phi)}[\phi_{+},\phi_{-}]&=\int\!\mathcal{D}\xi_0^{(\psi_h)}\;\mathbb{P}_{\psi_h}[\xi_0^{(\psi_h)}]\!\int\!\mathcal{D}\xi^{(\chi_{h})}\;\mathbb{P}_{\varsigma_{h}}[\xi^{(\varsigma_{h})}]\\
    &\qquad\qquad\qquad\times\exp\biggl\{i\,g_{\textsc{m}}^{2}\sum_{i,j}\int^t\!\!dt'\,dt''\;\Delta^{(\phi)}(\bm{a}_{i},t')\,G_{\mathrm{R}}^{(\varsigma_h)}(\bm{a}_i,t';\bm{a}_j,t'')\,\Sigma^{(\phi)}(\bm{a}_{j},t'')\biggr\}\notag\\
 \times\exp\biggl\{i\,g_{\textsc{m}}&\sum_i\int^t\!\!dt'\;\Delta^{(\phi)}(\bm{a}_{i},t')\,\Bigl[\xi^{(\varsigma_h)}(\bm{a}_i,t')+g_{\textsc{b}}\sum_{j}\int^{t'}\!\!dt''\;G_{\mathrm{R}}^{(\varsigma_h)}(\bm{a}_i,t';\bm{a}_j,t'')\,\xi_0^{(\psi_h)}(\bm{a}_j,t'')\Bigr]\biggr\}\,.\notag   
\end{align}	
Comparing with Eq.~\eqref{E:ifdghn},  we see  1) that the identification \eqref{E:wewsdw} is indeed valid, and 2) the appearance of an important identity
\begin{align}\label{E:iebera}
	&\quad\int\!\mathcal{D}\xi^{(\varsigma)}\;\mathbb{P}_{\varsigma}[\xi^{(\varsigma)}]\,\exp\biggl\{i\,g_{\textsc{m}}\sum_i\int^{t}\!\!dt'\;\Delta^{(\phi)}(\bm{a}_i,t')\,\xi^{(\varsigma)}(\bm{a}_i,t')\biggr\}\\
	&=\int\!\mathcal{D}\xi_{0}^{(\psi_{h})}\,\mathcal{D}\xi^{(\varsigma_{h})}\;\mathbb{P}_{\psi_{h}}[\xi_{0}^{(\psi_{h})}]\,\mathbb{P}_{\varsigma_{h}}[\xi^{(\varsigma_{h})}]\notag\\
	&\quad\times\exp\biggl\{i\,g_{\textsc{m}}\sum_i\int^{t}\!\!dt'\;\Delta^{(\phi)}(\bm{a}_i,t')\biggl[\xi^{(\varsigma_{h})}(\bm{a}_i,t')+g_{\textsc{b}}\sum_j\int^{t'}\!\!dt''\;G_{\mathrm{R}}^{(\varsigma_{h})}(\bm{a}_i,t';\bm{a}_j,t'')\,\xi_{0}^{(\psi_{h})}(\bm{a}_j,t'')\biggr]\biggr\}\,,\notag
\end{align}
which tells us in what context the identification \eqref{E:wewsdw} is meaningful. It also allows us to compute the covariance matrix element of the subsystem, here $\varsigma$,  completely within the framework of the stochastic effective action because it correctly incorporate the contribution of the homogeneous part.

{Before we move on to the effective actions of the ambient field modified by the dielectric medium, we give an example of the application of the identity~\eqref{E:iebera}. Suppose we would like to find the covariance matrix element $V_{\varsigma_i\varsigma_j}(t)$ for the subsystem $\varsigma$, defined by
\begin{equation}\label{E:wouss}
    V_{\varsigma_i\varsigma_j}(t)=\operatorname{Tr}_{\textsc{m}\textsc{b}}\biggl(\,\rho_{\textsc{m}}(t_i)\otimes\rho_{\textsc{b}}(t_i)\;\frac{1}{2}\bigl\{\hat{\varsigma}(\bm{a}_i,t),\,\hat{\varsigma}(\bm{a}_j,t)\bigr\}\biggr)\,,
\end{equation}
which is given by Eq.~\eqref{E:bverus} with $t'=t$. We can also obtain this by the identity~\eqref{E:iebera},
\begin{align}\label{E:ziutwe} 
    V_{\varsigma_i\varsigma_j}(t)&=\langle\hat{\varsigma}(\bm{a}_i,t)\,\hat{\varsigma}(\bm{a}_j,t)\rangle_{\xi^{(\varsigma)}}\\
    &=\biggl(\frac{1}{i\,g_{\textsc{m}}}\biggr)^2\frac{\delta^2}{\delta\Delta^{(\phi)}(\bm{a}_i,t)\,\delta\Delta^{(\phi)}(\bm{a}_j,t)}\int\!\mathcal{D}\xi^{(\varsigma)}\,\mathbb{P}_{\varsigma}[\xi^{(\varsigma)}]\,\exp\biggl\{i\,g_{\textsc{m}}\!\sum_i\!\int\!dt'\;\Delta^{(\phi)}(\bm{a}_i,t')\,\xi^{(\varsigma)}(\bm{a}_i,t')\biggr\}\notag\\
	&=\biggl(\frac{1}{i\,g_{\textsc{m}}}\biggr)^2\frac{\delta^2}{\delta\Delta^{(\phi)}(\bm{a}_i,t)\,\delta\Delta^{(\phi)}(\bm{a}_j,t)}\int\!\mathcal{D}\xi_{0}^{(\psi_{h})}\,\mathcal{D}\xi^{(\varsigma_{h})}\;\mathbb{P}_{\psi_{h}}[\xi_{0}^{(\psi_{h})}]\,\mathbb{P}_{\varsigma_{h}}[\xi^{(\varsigma_{h})}]\notag\\
	&\qquad\qquad\qquad\qquad\qquad\qquad\times\exp\biggl\{i\,g_{\textsc{m}}\sum_i\int^{t}\!\!dt'\;\Delta^{(\phi)}(\bm{a}_i,t')\biggl[\xi^{(\varsigma_{h})}(\bm{a}_i,t')\biggr.\biggr.\notag\\
        &\qquad\qquad\qquad\qquad\qquad\qquad\qquad\qquad\qquad+\biggl.\biggl.g_{\textsc{b}}\sum_j\int^{t'}\!\!dt''\;G_{\mathrm{R}}^{(\varsigma_{h})}(\bm{a}_i,t';\bm{a}_j,t'')\,\xi_{0}^{(\psi_{h})}(\bm{a}_j,t'')\biggr]\biggr\}\notag\\
        &=\langle\biggl[\xi^{(\varsigma_{h})}(\bm{a}_i,t)+g_{\textsc{b}}\sum_k\int^{t}\!\!dt'\;G_{\mathrm{R}}^{(\varsigma_{h})}(\bm{a}_i,t;\bm{a}_k,t')\,\xi_{0}^{(\psi_{h})}(\bm{a}_k,t')\biggr]\notag\\
        &\qquad\qquad\qquad\qquad\times\biggl[\xi^{(\varsigma_{h})}(\bm{a}_j,t)+g_{\textsc{b}}\sum_l\int^{t}\!\!dt''\;G_{\mathrm{R}}^{(\varsigma_{h})}(\bm{a}_j,t;\bm{a}_l,t'')\,\xi_{0}^{(\psi_{h})}(\bm{a}_l,t'')\biggr]\rangle_{\xi^{(\varsigma_{h})},\,\xi_{0}^{(\psi_{h})}}\notag\\
        &=\langle\xi^{(\varsigma_{h})}(\bm{a}_i,t)\xi^{(\varsigma_{h})}(\bm{a}_j,t)\rangle_{\xi^{(\varsigma_{h})}}\notag\\
        &\qquad+g_{\textsc{b}}^2\sum_{k,l}\int^t\!\!dt'\!\int^{t}\!\!dt''\;G_{\mathrm{R}}^{(\varsigma_h)}(\bm{a}_i,t;\bm{a}_k,t')\,G_{\mathrm{R}}^{(\varsigma_h)}(\bm{a}_j,t;\bm{a}_l,t'')\,\langle\xi_{0}^{(\psi_{h})}(\bm{a}_k,t')\,\xi_{0}^{(\psi_{h})}(\bm{a}_l,t'')\rangle_{\xi_{0}^{(\psi_{h})}}\,.\notag
\end{align}
Following Eqs.~\eqref{E:bwiys} and \eqref{E:qbkder}, we find that Eq.~\eqref{E:ziutwe} gives the same result as Eq.~\eqref{E:bverus} with $t'=t$, and thus is equivalent to Eq.~\eqref{E:wouss}.
}

\subsection{Key points in treating multi-partite open quantum systems, nested noises}
To summarize what we have done so far we use the derivations and results  obtained at this level of structure to identify the key points. This has broader meaning for the treatment of open multi-partite quantum systems under an ordered sequence of coarse-grainings, beyond the single system + environment cases a lot more commonly studied,  e.g., with the model of quantum Brownian motion. In particular, how to package the noises from the subsystems in the lower tiers into a more complete stochastic noise of the particular subsystem of interest.   

{
\begin{enumerate}
    \item In the conventional treatment of the Brownian motion, we only introduce the stochastic noise $\xi^{(\psi_h)}$ and its probability distribution functional $\mathbb{P}_{\psi_h}$ of the bath $\psi$ (translated to the notation here). We can derive a Langevin equation for the Brownian oscillator in a form like Eq.~\eqref{E:sltie} without the additional force from $\phi$. When we  compute the physical quantities, say, the covariance matrix elements of the Brownian oscillator $\varsigma$, we start from the complete solution like Eq.~\eqref{E:geres}. Then in this case we do not know how to compute the contribution from $\varsigma_h$ because we do not have the probability distribution functional $\mathbb{P}_{\varsigma_h}$ yet. Conventionally this quandary is circumvented by taking a hybrid approach, that is, viewing $\varsigma_h$ as an quantum operator so that its contribution can be found by the trace with respect to the initial state $\rho_{\textsc{m}}(t_i)$ of $\varsigma$. 
    \item Suppose we obtain $\mathbb{P}_{\varsigma_h}$, we can compute the right hand side of  Eq.~\eqref{E:geres} based on the knowledge of $\mathbb{P}_{\varsigma_h}$ and $\mathbb{P}_{\psi_h}$, but this may not guarantee a consistent result because we still do not know how to compute the left hand side of Eq.~\eqref{E:geres}, which is supposed to have its own probability distribution functional like $\mathbb{P}_{\varsigma}$.
    \item Finally, even though we obtain $\mathbb{P}_{\varsigma}$ by other means, there is no assurance of the correct relations  among $\mathbb{P}_{\varsigma}$, $\mathbb{P}_{\varsigma_h}$ and $\mathbb{P}_{\psi_h}$.
    \item Here, the identity Eq.~\eqref{E:iebera}   ensures the self-consistency among $\mathbb{P}_{\varsigma}$ and the $\mathbb{P}_{\varsigma_h}$, $\mathbb{P}_{\psi_h}$ pair, and allows us to naturally incorporate the contribution of $\varsigma_h$ completely within the stochastic formalism. {The example above explains how it works.}
    \item An important function of Eq.~\eqref{E:wewsdw} is that it guides the packaging of noises from the subsystems in the lower tiers into the (complete) stochastic noise of the particular subsystem of interest. The subsystem at hand need not be the one at the topmost level. Thus we can use this characteristic to develop a more compact or flexible formulation for a theory of multi-partite interacting open quantum systems.
    \item The derivations above show that we can work on the coarse-grained effective actions and the stochastic effective actions in parallel at any specific tier. We can transform one to the other and vice versa, and get a unique and correct answer within that tier which can enter into the dynamics of the variables of the reduced system in the next tier. 
    \item However, beware of confusion in the notations. Depending on how the stochastic noises from lower tiers are packaged, there could be more than one type of stochastic effective action or stochastic influence action. But they in principle can be differentiated by context and by noticing their origins.
    \item The influence actions are nothing but the Schwinger-Keldysh  or closed-time-path effective action in substance, an `in-in' completion of the traditional in-out effective action. Thus when additional external sources are included in the action, we can use the functional derivative with respect to the sources in a way similar to that in conventional field theory.
    \item By construction, from the stochastic effective action, taking the functional derivative only gives (the product of) the expectation values of the anti-commutators of the quantum operators of the subsystem. This may not pose any problem if the physical observables are required to take on real values, since their corresponding operators are typically arranged in symmetric operator ordering. If somehow we are interested in quantities of a different operator ordering, we can still find their expectation values from the coarse-grained effective action of the system via a similar procedure as illustrated here. They can be retrieved from the stochastic effective action by the stochastic average defined with respect to the appropriate probability distribution functionals.
\end{enumerate}
}

{Finally, a noteworthy point, observe that various expressions of the influence functional $W^{(\phi)}[\phi_+,\phi_-]$ share the same common factor 
\begin{equation*}
    g_{\textsc{m}}^{2}\sum_{i,j}\int^t\!\!dt'\,dt''\;\Delta^{(\phi)}(\bm{a}_{i},t')\,G_{\mathrm{R}}^{(\varsigma_h)}(\bm{a}_i,t';\bm{a}_j,t'')\,\Sigma^{(\phi)}(\bm{a}_{j},t'')\,.
\end{equation*}
Later we will see that this contribution associated with the retarded Green's function $G^{(\varsigma_h)}_{\mathrm{R}}$   quantifies  the response of the dielectric to the ambient field and can be identified as its dynamical susceptibility. This centerpiece quantity of many quantum theories of dielectrics is usually  derived in the context of linear response theory which relies on the existence of a thermal equilibrium condition. Here it will appear in the context of a fully nonequilibrium condition for the stochastic dynamics of multi-partite open quantum systems. 
}

\section{Effective Actions and Langevin Equation for the Ambient Field}
Continuing on in the same way as  in Eq.~\eqref{E:bfiers}, we write down a real stochastic effective action for the ambient field modified by the dielectric
\begin{align}
	&\hphantom{=}S_{\textsc{se}}^{(\phi)}[\phi_{+},\phi_{-};\xi^{(\varsigma_h)},\xi_0^{(\psi_h)}]\notag\\
    &=\int\!d^4x'\;\Bigl[-\frac{1}{2}\,\Delta^{(\phi)}(\bm{x}',t')\,\square_{x'}\Sigma^{(\phi)}(\bm{x}',t')-\frac{1}{2}\,\Sigma^{(\phi)}(\bm{x}',t')\,\square_{x'}\Delta^{(\phi)}(\bm{x}',t')\Bigr]\notag\\
    &\qquad\qquad\qquad\qquad\qquad\qquad+g_{\textsc{m}}^2\sum_{i,j}\int^{t}\!\!dt'\,dt''\;\Delta^{(\phi)}(\bm{a}_i,t')\,G_{\mathrm{R}}^{(\varsigma_h)}(\bm{a}_i,t';\bm{a}_j,t'')\,\Sigma^{(\phi)}(\bm{a}_j,t'')\notag\\
    &\qquad+g_{\textsc{m}}\sum_i\int^{t}\!\!dt'\;\Delta^{(\phi)}(\bm{a}_{i},t')\Bigl[\xi^{(\varsigma_h)}(\bm{a}_{i},t')+g_{\textsc{b}}\sum_j\int^{t'}\!\!d\tau'\;G_{\mathrm{R}}^{(\varsigma_h)}(\bm{a}_{i},t';\bm{a}_{j},\tau')\,\xi_0^{(\psi_h)}(\bm{a}_{j},\tau')\Bigr]\,,\label{E:sfkte}
\end{align}
with $\square_x=\partial_t^2-\bm{\nabla}_x^2$, from which we can derive a stochastic equation of motion for the dielectric-modified ambient field $\phi$.

Since
\begin{equation}
    \frac{\delta\phi(\bm{x}',t')}{\delta\phi(\bm{x},t)}=\delta^{(3)}(\bm{x}-\bm{x}')\,\delta(t-t')\,,
\end{equation}
the variation of the stochastic effective action with respect to $\Delta^{(\phi)}(\bm{x},t)$ gives the Langevin equation for the dielectric-modified quantum field:
\begin{align}
	\bigl(\partial_{t}^{2}-\bm{\nabla}_x^{2}\bigr)\phi(\bm{x},t)-g_{\textsc{m}}\sum_{i,j}\delta^{(3)}(\bm{x}-\bm{a}_i)&\int^{t}\!dt'\;G^{(\varsigma_h)}_{\mathrm{R}}(\bm{a}_i,t;\bm{a}_j,t')\,\Bigl[g_{\textsc{m}}\,\phi(\bm{a}_j,t')+g_{\textsc{b}}\,\xi_0^{(\psi_h)}(\bm{a}_j,t')\Bigr]\notag\\
	&=g_{\textsc{m}}\sum_i\delta^{(3)}(\bm{x}-\bm{a}_i)\,\xi^{(\varsigma_h)}(\bm{a}_i,t)\,.\label{E:oeiwjes}
\end{align}
It is instructive to decipher the physical meanings of each term in this equation of motion. The right hand side contains the intrinsic component of the dipole of the dielectric atom at  position $\bm{a}_i$. The delta function $\delta^{(3)}(\bm{x}-\bm{a}_i)$ restricts its contribution to within the dielectric. The nonlocal term on the left hand side is the induced dipole moment of the same atom at $\bm{a}_i$, often known as the polarization field. Note that this contribution is induced by both the full ambient field $\phi$ and the free thermal bath $\xi^{(\psi_h)}_0$. The ambient field involved must be the full field, not the free field, as is already known in  classical electromagnetism, to ensure self-consistency of the whole enterprise.  The part involving the dipoles induced by the ambient field can be conversely understood as the backreaction  of the dipole on the ambient field. This nonlocal term is connected with the electric susceptibility, as can be seen more readily with a Fourier transformation. We can write the interacting wave equation \eqref{E:oeiwjes} in the frequency domain as
\begin{align}\label{E:qyrwre}
    \bm{\nabla}_x^{2}\tilde{\phi}&(\bm{x},\omega)+\omega^2\tilde{\phi}(\bm{x},\omega)+g_{\textsc{m}}^2\sum_{i,j}\delta^{(3)}(\bm{x}-\bm{a}_i)\,\tilde{G}^{(\varsigma_h)}_{\mathrm{R}}(\bm{a}_i,\bm{a}_j;\omega)\,\tilde{\phi}(\bm{a}_j,\omega)\\
    &=-g_{\textsc{m}}\sum_i\delta^{(3)}(\bm{x}-\bm{a}_i)\,\tilde{\xi}^{(\varsigma_h)}(\bm{a}_i,\omega)-g_{\textsc{m}}g_{\textsc{b}}\sum_{i,j}\delta^{(3)}(\bm{x}-\bm{a}_i)\,\tilde{G}^{(\varsigma_h)}_{\mathrm{R}}(\bm{a}_i,\bm{a}_j;\omega)\,\tilde{\xi}_0^{(\psi_h)}(\bm{a}_j,\omega)\,,\notag
\end{align}
where the tilde above a quantity denotes its Fourier transformation, defined by
\begin{equation}
    f(t)=\int\!\frac{d\omega}{\sqrt{2\pi}}\;\tilde{f}(\omega)\,e^{-i\omega t}\,.
\end{equation}
Note that the formalism we adopted and implemented is valid for fully nonequilibrium dynamics, while Fourier representation presumes a stationary configuration. Thus before invoking these expressions one needs to justify the assumption of an equilibrium condition.

\subsection{Connect with macro theories: Lorentz model of the dielectric medium}
We will further write Eq.~\eqref{E:qyrwre} into a more familiar form
\begin{align}\label{E:eruhes}
    \bm{\nabla}_x^{2}\tilde{\phi}(\bm{x},\omega)+\omega^2\int\!d^3\bm{y}\;\tilde{\varepsilon}(\bm{x},\bm{y};\omega)\,\tilde{\phi}(\bm{y},\omega)=-\tilde{\varrho}(\bm{x},\omega)\,,
\end{align}
where we have introduced the spatially inhomogeneous dielectric function $\tilde{\varepsilon}(\bm{x},\bm{y};\omega)$ in the frequency domain
\begin{equation}
    \tilde{\varepsilon}(\bm{x},\bm{y};\omega)=\delta^{(3)}(\bm{x}-\bm{y})+\frac{g_{\textsc{m}}^2}{\omega^2}\sum_{i,j}\delta^{(3)}(\bm{x}-\bm{a}_i)\,\tilde{G}^{(\varsigma_h)}_{\mathrm{R}}(\bm{a}_i,\bm{a}_j;\omega)\,\delta^{(3)}(\bm{a}_j-\bm{y})\,,
\end{equation}
and the bounded polarizable source $\tilde{\varrho}(\bm{x},\omega)$
\begin{equation}\label{E:edbkdfgd}
    \tilde{\varrho}(\bm{x},\omega)=g_{\textsc{m}}\sum_i\delta^{(3)}(\bm{x}-\bm{a}_i)\,\tilde{\xi}^{(\varsigma_h)}(\bm{a}_i,\omega)+g_{\textsc{m}}g_{\textsc{b}}\sum_{i,j}\delta^{(3)}(\bm{x}-\bm{a}_i)\,\tilde{G}^{(\varsigma_h)}_{\mathrm{R}}(\bm{a}_i,\bm{a}_j;\omega)\,\tilde{\xi}_0^{(\psi_h)}(\bm{a}_j,\omega)\,.
\end{equation}
This bounded polarizable source $\varrho$ includes only the intrinsic component of the dipole moments of the dielectric atoms and the contribution induced by the bath. It does not contain the dipole moments induced by the ambient field. The latter enters in the susceptibility. The microscopic spatially inhomogeneous electric susceptibility $\chi_e(\bm{x},\bm{y};\omega)$ is then given by
\begin{equation}
    \chi_e(\bm{x},\bm{y};\omega)=\frac{g_{\textsc{m}}^2}{\omega^2}\sum_{i,j}\delta^{(3)}(\bm{x}-\bm{a}_i)\,\tilde{G}^{(\varsigma_h)}_{\mathrm{R}}(\bm{a}_i,\bm{a}_j;\omega)\,\delta^{(3)}(\bm{a}_j-\bm{y})\,.
\end{equation}
The role of the retarded Green's function of the dipole moment of the dielectric atom is clear. These results are refinements of earlier results presented in Eq.~(36) of \cite{BH11}. The retardation or the causal property of the susceptibility function naturally and beautifully emerges in this formalism without being a priori imposed. If we ignore the mutual causal influences between the dielectric atoms, then the retarded Green's function given by Eq.~\eqref{E:aesobde} has the form in frequency space: 
\begin{equation}\label{E:nldjngd}
    \frac{1}{m(-\omega^2-i\,2\gamma_{\textsc{b}}\omega+\varpi_{\textsc{r}}^2)}\,.
\end{equation}
This is exactly the expression in the Lorentz model of the dielectric in classical electromagnetic theory~\cite{Ja98}. Otherwise, the denominator in \eqref{E:nldjngd} will contain contributions from the mutual influences proportional to $\tilde{G}_{\mathrm{R},0}^{(\psi_h)}(\bm{a}_i,\bm{a}_j;\omega)$. We also note that the bounded polarizable source $\varrho(\bm{x},t)$ is directly related to $\xi^{(\varsigma)}(\bm{a}_i,t)$ in Eq.~\eqref{E:wewsdw} by
\begin{equation}
    \varrho(\bm{x},t)=g_{\textsc{m}}\sum_i\delta^{(3)}(\bm{x}-\bm{a}_i)\,\xi^{(\varsigma)}(\bm{a}_i,t)\,.
\end{equation}

Eq.~\eqref{E:eruhes} gives the nonequilibrium microscopic description of the ambient field under the influence of the quantum dielectric. It applies to the regions  {inside and outside} the dielectric, as can be seen by the presence of the delta functions. To obtain a  macroscopic description of the ambient field modified by the dielectric one needs to introduce suitable spatial and/or temporal averaging procedures~\cite{Ja98,Ru70,Ro71,Ro73}. The averaging scales will delimit the ranges of validity of any such macroscopic theory.

\subsection{Stochastic bound sources are needed to ensure equal-time commutation relations}
Eq.~\eqref{E:eruhes} can be understood as the scalar counterpart of the electromagnetic field wave equation in the presence of the dielectric medium (apart from the averaging protocols). The nice things about the modified equation of scalar field \eqref{E:eruhes} are that 1) its form bears intuitive resemblance to the classical wave equation of the field in the presence of the dielectric and 2) the net effects of the quantum linear dielectric are neatly summarized in the dielectric function $\tilde{\varepsilon}$ and the stochastic bound source $\tilde{\varrho}$. Therefore our knowledge regarding the interaction between the classical (electromagnetic) field and the classical dielectric can be readily applied here. For example, we thus expect that for the field inside the dielectric body, its amplitude and phase will be modified by the dielectric function, and further altered by the quantum fluctuations associated with the dielectric. Notice that the bounded polarizable source $\tilde{\varrho}$,  defined in \eqref{E:edbkdfgd}, results from the quantum fluctuations of the intrinsic dipoles of the constituent atoms of the dielectric and the induced dipoles from the thermal bath. It is an aggregation of point-like sources, located at the position of each constituent atom. Hence the field outside the dielectric will follow a homogeneous wave equation. In addition, these point-like sources in the dielectric is an idealization of the real atom if  we are interested in phenomena at a scale larger than the size of the atom in the dielectric. Oftentimes   some averaging procedures are introduced to smooth out the spatial distribution of the atoms, so that when viewed from a distance, the dielectric has a spatially continuous, smooth or even uniform distribution of the bounded sources. Here we stress that the presence of the stochastic sources, though not present in classical theory, is necessary in order to guardrail the equal-time commutation relations among the canonical variables of each subsystems. Without them, the commutation relations can decay to vanish, violating unitarity. In our formalism, validity of the equal-time commutation relations are automatically ensured and we do not need to impose them a priori.

Finally we note that although the fluctuation and dissipation kernels are nonlocal in time, the obtained Maxwell-Langevin equation remains an exact description of the dynamics of the field~\cite{CRV}. By this we mean that the state of the field follows exactly from taking the average over the initial conditions and the evolution trajectories driven by the stochastic source. Furthermore, a subclass of quantum correlation functions, including the field's Hadamard Green's function can also be deduced from this description.

Hereafter we will assume that a suitable spatial averaging process has been carried out to arrive at the macroscopic counterpart of Eq.~\eqref{E:eruhes}, such that the dielectric function becomes spatially-homogeneous. This facilitates a smooth connection with the macroscopic theory of quantized (electromagnetic) field, where the properties of the field are primarily determined by the dielectric function. In terms of practical applications, on  scales much greater than the inter-atomic distances, we only see the coarse-grained manifestations of the dielectric material, regardless of its underlying atomic/molecular details. At this coarse-grained level of accuracy one can use the empirical data of the medium from  experiments to  construct a phenomenological model for the dielectric functions~\cite{Ch10,Sa19,Ro73,Bu12}. This procedure can be implemented using the results we have obtained: The relevant retarded Green's function of the $\phi$ field will satisfy an inhomogeneous wave equation with a delta function source,
\begin{equation}\label{E:dkfbws}
	\Bigl[\bm {\nabla}_x^{2}+\omega^{2}\tilde{\varepsilon}\bigl(\bm{x},\omega\bigr)\Bigr]\tilde{G}_{\mathrm{R}}^{(\phi)}(\bm{x},\bm{x}';\omega)=-\delta^{(3)}(\bm{x}-\bm{x}')\,,
\end{equation}
under appropriate boundary conditions, while the Hadamard function is the homogeneous solution of the same wave function,
\begin{equation}\label{E:wjdfhbws}
	\Bigl[\bm{\nabla}^{2}+\omega^{2}\tilde{\varepsilon}\bigl(\bm{x},\omega\bigr)\Bigr]\tilde{G}_{\mathrm{H}}^{(\phi)}(\bm{x},\bm{x}';\omega)=0\,.
\end{equation}
The dielectric function $\varepsilon$ can either be derived from first principles using a microphysics model as described here, or determined by experimental data as nicely described in \cite{Bu12}. {The spatial dependence in the dielectric function $\tilde{\varepsilon}\bigl(\bm{x},\omega\bigr)$ in Eqs.~\eqref{E:dkfbws} and \eqref{E:wjdfhbws} merely serves to indicate the boundary interface of the bulk dielectric.}

We now proceed to calculate the influence actions of this medium-altered ambient quantum field on the $N$ neutral atoms  {residing in the field}.

\section{Stochastic Effective Action and Langevin Equation for the Dipoles of \texorpdfstring{$N$}{N} Neutral Atoms}
We are now in the last step, namely, integrating over the dielectric-modified field variable to see how  they affects the dynamics of the internal degrees of freedom $\chi^{(n)}$ of the $N$ atoms, at fixed location $\bm{z}_{n}$ outside the bulk medium, with $n=1,\,\ldots,\, N$. Since by assumption all of these are linearly coupled to the scalar field, the influence of the field on the reduced system can be fully described by an influence action. From \eqref{E:djfek} and \eqref{E:sfkte}, we note that the influence functional on the dipoles of $N$ neutral atoms, $\mathcal{W}^{(\chi)}[\{\chi_{+}\},\{\chi_{-}\}]$ is formally given by
\begin{align}\label{E:ernks}
	\mathcal{W}^{(\chi)}[\{\chi_{+}\},\{\chi_{-}\}]&=\int\!d\phi_{f}\int_{\phi_{i}}^{\phi_{f}}\!\mathcal{D}\phi_{+}\!\int_{\phi'_{i}}^{\phi_{f}}\!\mathcal{D}\phi_{-}\,\exp\biggl\{i\,S_{\textsc{f}}[\phi_{+}]-i\,S_{\textsc{f}}[\phi_{-}]+i\,S_{\textsc{sif}}^{(\phi)}[\phi_{+},\phi_{-};\xi^{(\varsigma_h)},\xi_0^{(\psi_h)}]\biggr.\notag\\
	&\qquad\qquad\qquad\qquad\qquad+\biggl.i\,S_{\textsc{af}}[\{\chi_{+}\},\phi_{+}]-i\,S_{\textsc{af}}[\{\chi_{-}\},\phi_{-}]\biggr\}\circ\rho_{\textsc{f}}(t_{i})\,,
\end{align}
where $\{\chi\}$ is a shorthand notation for the collection $(\chi^{(1)}$, $\chi^{(2)}$, $\dots$, $\chi^{(n)})$, i.e., the displacements of the internal degree freedoms of $N$ neutral atoms. This influence functional, though similar to \eqref{E:vbxir}  {in structure}, already contains the influences from the dielectric, as shown in an additional interaction term in $S_{\textsc{sif}}^{(\phi)}$ that describes the coupling between the field and the stochastic sources from the medium. By comparing with the earlier results \eqref{E:vbxir} and \eqref{E:reuthsiud}, we can readily deduce the influence functional $S_{\textsc{sif}}^{(\chi)}$ for the $N$ atoms if we observe that $\varrho(\bm{x},t)$ is nothing but the driving sources on the right hand side of wave equation of the medium-modified field~\eqref{E:eruhes}.

{Following this line of arguments, we first} write $S_{\textsc{se}}^{(\phi)}$ in Eq.~\eqref{E:sfkte} into
\begin{align}
    &\hphantom{=}S_{\textsc{se}}^{(\phi)}[\phi_{+},\phi_{-};\xi^{\varsigma_h},\xi_0^{\psi_h}]\notag\\
    &=\int^x\!\!d^4x'\;\Bigl[-\frac{1}{2}\,\Delta^{(\phi)}(\bm{x}',t')\,\square_{x'}\Sigma^{(\phi)}(\bm{x}',t')-\frac{1}{2}\,\Sigma^{(\phi)}(\bm{x}',t')\,\square_{x'}\Delta^{(\phi)}(\bm{x}',t')\Bigr]\notag\\
    &\qquad\quad+g_{\textsc{m}}^2\int^{x}\!\!d^4x'\,d^4x''\;\Delta^{(\phi)}(\bm{x}',t')\Bigl[\sum_{i,j}\delta^{(3)}(\bm{x}'-\bm{a}_i)\,G_{\mathrm{R}}^{(\varsigma_h)}(\bm{a}_i,t';\bm{a}_j,t'')\,\delta^{(3)}(\bm{x}''-\bm{a}_j)\Bigr]\Sigma^{(\phi)}(\bm{x}'',t'')\notag\\
    &\qquad\qquad\int^x\!\!d^4x'\;\Sigma^{(\phi)}(\bm{x}',t')\,\Delta^{(J^{(\chi)})}(\bm{x}',t')+\int\!d^4x'\;\Delta^{(\phi)}(\bm{x}',t')\,\Sigma^{(J^{(\chi)})}(\bm{x}',t')\,,
\end{align}
where $J_{\pm}^{(\chi)}(\bm{x},t)=\displaystyle g_{\textsc{f}}\sum_{n}\chi_{\pm}^{(n)}(t)\,\delta^{(3)}(\bm{x}-\bm{z}_{n})+\varrho(\bm{x},t)$ that account for the totality of the dipole moments of $N$ atoms and those contributions from the dielectric atoms, such that
\begin{align}
    \Sigma^{(J^{(\chi)})}(\bm{x},t)&=g_{\textsc{f}}\sum_{n}\Sigma^{(\chi^{(n)})}(t)\,\delta^{(3)}(\bm{x}-\bm{z}_{n})+\varrho(\bm{x},t)\,,\\
    \Delta^{(J^{(\chi)})}(\bm{x},t)&=g_{\textsc{f}}\sum_{n}\Delta^{(\chi^{(n)})}(t)\,\delta^{(3)}(\bm{x}-\bm{z}_{n})\,.
\end{align}
{This is analogous to Eq.~\eqref{E:ljbieur}. Comparison with Eq.~\eqref{E:reuthsiud} implies that the influence action $S_{\textsc{sif}}^{(\chi)}$ for the $N$ atoms can be identified as}
\begin{align}
    S_{\textsc{sif}}^{(\chi)}[\{\chi_+\},\{\chi_-\};\varrho]&=\int\!d^4x'\,d^4x''\;\biggl[\Delta^{(J^{(\chi)})}(\bm{x}',t')\,G_{\mathrm{R}}^{(\phi_h)}(\bm{x}',t';\bm{x}'',t'')\,\Sigma^{(J^{(\chi)})}(\bm{x}'',t'')\biggr.\notag\\
    &\qquad\qquad\qquad\qquad+\biggl.\frac{i}{2}\,\Delta^{(J^{(\chi)})}(\bm{x}',t')\,G_{\mathrm{H}}^{(\phi_h)}(\bm{x}',t';\bm{x}'',t'')\,\Delta^{(J^{(\chi)})}(\bm{x}'',t'')\biggr]\,,
\end{align}
Note at this theoretical level  $\varrho(\bm{x},t)$ is not a dynamical variable but treated as an external variable. It is nonzero only inside the dielectric, and vanishes outside the medium.  {(Remember $\varrho(\bm{x},t)$ contains the stochastic noises $\xi^{(\varsigma_h)}$ and $\xi_0^{(\psi_h)}$.)}

The Green's functions $G_{\mathrm{R}}^{(\phi_h)}$ and $G_{\mathrm{H}}^{(\phi_h)}$ of the dielectric-modified scalar field have been introduced earlier, and satisfy the Eqs.~\eqref{E:dkfbws} and \eqref{E:wjdfhbws} if hereafter we are interested in only the macro-field. Similar to the earlier cases, the field used to construct these two Green's function come from the homogeneous solution of the modified wave equation, excluding the contributions from the bounded polarizable sources.

The stochastic effective action of the $N$ atoms $S^{(\chi)}_{\textsc{se}}[\{\chi_+\},\{\chi_-\};\varrho]=S_{\textsc{a}}[\{\chi_{+}\}]-S_{\textsc{a}}[\{\chi_{-}\}]+S_{\textsc{sif}}^{(\chi)}[\{\chi_{+}\},\{\chi_{-}\};\varrho]$ is then given by ,
\begin{align}\label{E:bdfery}
	S^{(\chi)}_{\textsc{se}}[\{\chi_+\},\{\chi_-\};\varrho]&=M\sum_{n}\int^t\!\!dt'\;\Bigl[\dot{\Delta}^{(\chi^{(n)})}(t')\,\dot{\Sigma}^{(\chi^{(n)})}(t')-\Omega^{2}\Delta^{(\chi^{(n)})}(t')\,\Sigma^{(\chi^{(n)})}(t')\Bigr]\\
	+\int\!d^{4}x'\,d^{4}x''&\;\biggl[\Delta^{(J^{(\chi)})}(x')\,G_{\mathrm{R}}^{(\phi_h)}(x';x'')\,\Sigma^{(J^{(\chi)})}(x'')+\frac{i}{2}\,\Delta^{(J^{(\chi)})}(x')\,G_{\mathrm{H}}^{(\phi_h)}(x';x'')\,\Delta^{(J^{(\chi)})}(x'')\biggr]\,.\notag
\end{align}
From this we readily {obtain one another stochastic effective action of the $N$ atoms by applying the Feynman-Vernon identity to the term containing $G_{\mathrm{H}}^{(\phi_h)}$. Thus we arrive at}
\begin{align}\label{E:lofbkd}
	S^{(\chi)}_{\textsc{se}}[\{\chi_+\},\{\chi_-\};\xi^{(\phi_h)},\varrho]&=M\sum_{n}\int^t\!\!dt\;\Bigl[\dot{\Delta}^{(\chi^{(n)})}(t')\,\dot{\Sigma}^{(\chi^{(n)})}(t')-\Omega^{2}\Delta^{(\chi^{(n)})}(t')\,\Sigma^{(\chi^{(n)})}(t')\Bigr]\\
	\int\!d^{4}x'\,&d^{4}x''\;\Delta^{(J^{(\chi)})}(x')\,G_{\mathrm{R}}^{(\phi_h)}(x';x'')\,\Sigma^{(J^{(\chi)})}(x'')+\int\!d^{4}x'\;\Delta^{(J^{(\chi)})}(x')\,\xi^{(\phi_h)}(x')\,,\notag
\end{align}
with the introduction of the probability distribution functional $\mathbb{P}_{\phi_h}[\xi^{(\phi_h)}]$ for the ensemble average with respect to the stochastic source $\xi^{(\phi_h)}(x)$, in a way similar to Eqs.~\eqref{E:bfiers} and~\eqref{E:sfkte}.  {Understandably this stochastic effective action has one more noise than the stochastic effective action in  Eq.~\eqref{E:bdfery}.} {This is an example that for a subsystem in the higher tiers, it can have more than one stochastic effective action, depending on how the stochastic noises from the subsystems in the lower tiers are packaged via an identity like~\eqref{E:iebera} together with an assignment similar to \eqref{E:wewsdw}.}

Its variation with respect to $\Delta^{(\chi^{(n)})}$ will give the Langevin equation for the nonequilibrium stochastic dynamics of the dipole in the $n^{\text{th}}$ neutral atom positioned at $\bm{z}_n$,
\begin{align}\label{E:eerbesds}
 	M\ddot{\chi}^{(n)}(t)+M\Omega^{2}\chi^{(n)}(t)&-g_{\textsc{f}}^{2}\sum_{n'}\int_{0}^{t}\!dt'\;G_{\mathrm{R}}^{(\phi_h)}(\bm{z}_{n},t;\bm{z}_{n'},t')\,\chi^{(n')}(t')\\
	&=g_{\textsc{f}}\,\xi^{(\phi_h)}(\bm{z}_{n},t)+g_{\textsc{f}}\int^{x}\!\!d^4x'\;G_{\mathrm{R}}^{(\phi_h)}(\bm{z}_{n},t;\bm{x}',t')\,\varrho(\bm{x}',t')\,.\notag
\end{align}
Similar to the discussion in Sec.~\ref{S:aket}, the nonlocal term on the left hand side includes more than the self-force associated with the dipole of the $n^{\text{th}}$ atom. It also contains the retarded effects from the dipoles of the other ($N-1$) atoms. If we place these $N$ atoms outside the dielectric, then from the macro-field viewpoint, owing to the existence of an interface between the vacuum and the dielectric medium, there will be additional retarded contributions from the image dipoles of all $N$ atoms.

The nonlocal expression on the right hand side results from the retarded action from the constituent atoms of the dielectric. However, its presence is not surprising from the perspective that the ambient scalar field is coupled to both the dielectric atoms and the $N$ atoms outside the medium. Hence there should be an indirect interaction between them, mediated by the ambient field. In a sense, it could be understood as the consequence that the stochastic polarizable sources $\varrho(\bm{x},t)$ in the dielectric emits some Li\'enard-Wiechert-like radiation, which transmits through the dielectric boundary to the location of the $n^{\text{th}}$ neutral atom, and interact with the dipole moment of the $n^{\text{th}}$ atom which also interacts with the on-site ambient scalar field.

The new noise force $\xi^{(\phi_h)}$ on the right hand side, which manifests the quantum fluctuations of the dielectric-modified scalar field in its initial state, satisfies a Gaussian statistics, such that
\begin{align}
	\langle\xi^{(\phi_h)}(x)\rangle_{\xi^{(\phi_h)}}&=0\,,&\langle\xi^{(\phi_h)}(x)\xi^{(\phi_h)}(x')\rangle_{\xi^{(\phi_h)}}&=G_{\mathrm{H}}^{(\phi_h)}(x;x')\,.
\end{align}
{As discussed earlier,} it corresponds to the homogeneous solution of the wave equation of the dielectric-modified field \eqref{E:oeiwjes}.

As has been mentioned earlier, we can recover the coarse-grained effective action $S_{\textsc{cgea}}^{(\chi)}[\{\chi_+\},\{\chi_-\}]$ of the $N$-atom subsystem from the stochastic effective action Eq.~\eqref{E:bdfery} or \eqref{E:lofbkd}. Taking \eqref{E:lofbkd} as an example, we can use
\begin{align}\label{E:djfkdw}
    &\hphantom{=}\exp\biggl\{i\,S_{\textsc{cgea}}^{(\chi)}[\{\chi_+\},\{\chi_-\}]\biggr\}\\
    &=\int\!\mathcal{D}\xi^{(\phi_h)}\,\mathbb{P}_{\phi_h}[\xi^{(\phi_h)}]\!\int\!\mathcal{D}\xi^{(\varsigma_h)}\,\mathbb{P}_{\varsigma_h}[\xi^{(\varsigma_h)}]\!\int\!\mathcal{D}\xi_0^{(\psi_h)}\,\mathbb{P}_{\psi_h}[\xi_0^{(\psi_h)}]\;\exp\biggl\{i\,S^{(\chi)}_{\textsc{se}}[\{\chi_+\},\{\chi_-\};\xi^{(\phi_h)},\varrho]\biggr\}\notag
\end{align}
to find $S_{\textsc{cgea}}^{(\chi)}[\{\chi_+\},\{\chi_-\}]$. The multiple path integrations can be carried out straightforwardly because the integrand is a Gaussian. In principle, we can use the left hand side of Eq.~\eqref{E:djfkdw} to obtain the elements of the reduced density matrix for the $N$-atom system,
\begin{align}
    \rho_{\textsc{a}}(\{\chi^{(n)}_f\},\{\chi'^{(n)}_f\},t)=\int_{\chi^{(n)}_i}^{\chi^{(n)}_f}\!\!\mathcal{D}\!\chi^{(n)}_+\int_{\chi'^{(n)}_i}^{\chi'^{(n)}_f}\!\!\mathcal{D}\!\chi^{(n)}_-\;\exp\biggl\{i\,S_{\textsc{cgea}}^{(\chi)}[\{\chi_+\},\{\chi_-\}]\biggr\}\circ\rho_{\textsc{a}}(t_i)\,.
\end{align}
This links back to Eq.~\eqref{E:dlied}. Then we can use Eq.~\eqref{E:dheruhd} to compute the physical observables for the $N$-atom system. Although theoretically it is nothing but a few more Gaussian integrations to derive the elements of the reduced density matrix operator $\rho_{\textsc{a}}$, in practice, it is simpler to use the trick of the functional derivative. Essentially in the same way as we did earlier to derive the influence functional $\mathcal{W}$ for each subsystem, we first find the in-in (closed-time-path) effective action for the $N$ atom in the presence of external sources $J^{(n)}_{\pm}$: 
\begin{align}
    \mathcal{W}[J^{(n)}_+,J^{(n)}_-]&=\int\!d\chi^{(n)}_f\!\int_{\chi^{(n)}_i}^{\chi^{(n)}_f}\!\!\mathcal{D}\!\chi^{(n)}_+\int_{\chi'^{(n)}_i}^{\chi^{(n)}_f}\!\!\mathcal{D}\!\chi^{(n)}_-\;\exp\biggl\{i\,S_{\textsc{cgea}}^{(\chi)}[\{\chi_+\},\{\chi_-\}]\biggr.\notag\\
    &\qquad\qquad\qquad\qquad\qquad+\biggl.\int\!dt'\;\Bigl[J^{(n)}_+(t')\,\chi^{(n)}_+(t')-J^{(n)}_-(t')\,\chi^{(n)}_-(t')\Bigr]\biggr\}\circ\rho_{\textsc{a}}(t_i)\,.
\end{align}
Then take the functional derivatives of $ \mathcal{W}[J^{(n)}_+,J^{(n)}_-]$ with respect to $J^{(n)}_{+}$, $J^{(n)}_{-}$ or $\Sigma^{J^{(n)}}$, $\Delta^{J^{(n)}}$. This will serve the same purpose. This functional derivative approach is particularly useful for perturbative treatments of systems with nonlinearity \cite{HPZ93,HH20a,HH20b,YH20,SLS,But2Mir}.

At the end of this discussion, we want to raise two noteworthy points -- 

1) on the quantum nature of the stochastic equations derived by taking the functional variation of the stochastic effective actions of the corresponding subsystems:  A clear benefit is that the Langevin equations obtained this way can provide better insights into the physical meanings of the nested expressions from the successive layers of effective actions. On the other hand, once one turns a quantum operator expression to a classical stochastic source such as invoking the Feynman-Vernon Gaussian identity these Langevin equations are $c$-number stochastic equations, so it is a legitimate concern as to what degree one can preserve the quantum nature of the operators in the subsystems. {Observe that if the physical observable is defined by the expectation value  of the product of two quantum Hermitian operators of a given subsystem, then these two operators should be arranged in a symmetric ordering or in the form of an anti-commutator, so that the corresponding expectation value  remains  real. {Now switch from the quantum view to the stochastic view.} By construction, the probability distribution functional is defined in terms of the Hadamard function of the operator  of the subsystem variable, which is the expectation value of its anti-commutator.  Thus, if the quantity is given by a product of the subsystem's variables, i.e., the solutions of this $c$-number Langevin equation of the subsystem, then its {statistical ensemble average over noise} will be equivalent to the expectation value of the corresponding quantum operators {\it in symmetric ordering}.} In Paper II we shall show a more direct way to derive the quantum Langevin equations for the same setup of quantum systems and their interactions.

2) on the difference between taking the coarse-grained versus the stochastic effective action route:  If we are only interested in finding the reduced density matrix elements of the internal degrees of freedom of the $N$-atom subsystem, which sits at the highest (outermost) level of the nested set, we do not need to derive the stochastic effective actions of the subsystems in lower levels. We just integrate over all variables of the subsystems, obtaining various coarse-grained effective actions one after another. This route, though in appearance looks monstrously complicated { and physically less transparent}, will reveal its true power {1) when one is interested in the cases when the product of operators is arranged in ordering other than the symmetric one, or 2) }when nonlinearity emerges in the processes of tracing over the subsystems. If perturbative treatment for finding solution is warranted, many functional integral techniques in field theory can be readily applied~\cite{HPZ93,HH20a,HH20b,YH20,SLS,But2Mir}.  In contrast, it is usually much more difficult to handle the  {corresponding} operator equations perturbatively.

\section{Summary and Outlook}
To summarize, we  take a broader perspective in what this first paper  has accomplished in relation to the important existing works,  and what new prospects it holds for the study of AFM interactions in future investigations. Three primary features of this paper are: 1) Further developments of the influence functional formalism for the treatment of {\it multi-partite open quantum systems}; 2) {\it Microphysics modeling of the dielectric medium}. We use the simplest micro-physics model as example and build up to the macroscopic level, connecting with the Lorentz model of dielectrics; 3) A new paradigm for detailed systematic studies of the {\it nonequilibrium stochastic dynamics} of atoms interacting with a quantum field in the presence of a medium, including self-consistent backactions at all levels.

{\it Microphysics modeling of the Medium}.   We mention 2) first,  because the dielectric medium is where much effort in AFM interaction were invested in and the most familiar to readers, but carries the least weight in our design and contributions. We identified two important models, the Hutner Barnett (HB) \cite{HB} model based on microphysics where the medium and its bath are treated as a closed system. We also used a microphysics model, the ubiquitous harmonic oscillators, but our treatment, same as earlier work by Behunin and Hu (BH) \cite{BH11},   is by way of open quantum systems. So it would be of some theoretical interest to compare the two approaches.  The other major theory in the trade is called macroscopic QED or stochastic electrodynamics, which was started by Lifshitz, developed by Pitaevsky, Welsch and their co-authors, et al and expounded in great detail in the books of Buhmann \cite{Bu12,Bu13}. This popular model is likely of more practical use as it inputs experimental data to fix the parameters in the medium's dynamical susceptibility functions. It would be of interest to see how a microphysics model constructed from first principles, such as in HB, BH and this work, can provide better justifications to the basic premises used in the semi-phenomenological models, such as causality in nonMarkovian quantum dynamics,  positivity in stochastic dynamics and  self-consistency in the treatment of back-actions, instead of positing them as requisite conditions.

{\it Systematics for Multipartites}:  For 1) we have three layers of structures and four (or five, if the external dof of the atom is included) variables, two for the dielectric, one for the field, one (or two) for the atoms. Traditional treatment  of open systems has one system variable and one (set) for the bath. It is clear for bipartites how the bath influences the system, a well-known example is quantum Brownian motion. But how would the noise of the bath in the medium show up in the quantum field and in the system atoms?  Do we see some systematics in the hierarchy of noises entering in the stochastic equations of motion for each level of structure after successive coarse-grainings? Indeed there are: we have derived expressions showing how noises at each sublevel of structure enter at a higher level. For details, please refer to Sec.~\ref{S:aket}. We have also clarified the role of the complex coarse-grained effective action (before noise is identified) and the real stochastic effective action (noise appears a la Feynman-Vernon identity) in an orderly succession of coarse-grainings, explaining the rich meaning of `graded' influence action in the influence functional formalism. Finally we have also answered the question concerning the degree and extent a stochastic representation of a multi-partite quantum system can retain the quantumness of that system. Basically, for Gaussian systems, in full, if noise appears by way of the FV identity, but not necessarily so, especially when put in by hand, without self-consistency considerations. 

{\it Prospect and Follow-ups}:  With these niceties, for 3),  the systematics for treating multi-partite open  quantum systems begun in \cite{BH11} and further advanced in this work can be used for treating quantum informational issues such as the entanglement dynamics in AFM interactions. One can calculate the  covariant matrix elements of the correlation functions of the idf of $N$-atoms in a dielectric-altered quantum field. The hierarchical ordering of physical parameters resulting from coarse-graining the sublevel variables gives a clear physical picture of their inter-level dependence. This calculation is undertaken in Paper II. Exact analytic results can be obtained for the combined system at late times when all subsystems settle into equilibrium. As a  useful illustration we consider just one atom and use the covariance matrix elements to calculate the purity of the system interacting with a quantum field in the  presence of a  dielectric half plane.  In Paper III we use these results to investigate the entanglement between an atom and a dielectric half space. We derive the entanglement domain on the dielectric and  contrast these results with prior findings of the entanglement between an atom in the presence of a dielectric \cite{Klich}, a conductor \cite{Rong} and in free space.  

In subsequent work the external or mechanical degree of freedom (edf or mdf) for the atom will be invoked \cite{MovA1,MovA2}, in joint force with the work based on the mirror-oscillator-field (MOF) model \cite{GBH,SLH,LinMOF,SLS,But2Mir,ButN-D} of quantum optomechanics mentioned earlier for moving mirror-quantum field interactions. This confluence  completes the theoretical basis of  the graded influence action formalism for multi-partite open quantum systems in  AFM interactions, capable of describing moving atoms or moving mirrors interacting with a dielectric-altered quantum field. This enterprise will enable one to  tackle many interesting effects of fundamental physics values, from the dynamical Casimir and dynamical Casimir-Polder effects to quantum friction, and, in the relativistic realm, Hawking effect \cite{Haw75} and its moving mirror analogues \cite{DavFul,DavFul1},  Unruh effect \cite{Unr76},  and acceleration radiation \cite{Scully}.

\quad\\
\noindent{\bf Acknowledgments} 
 J.-T. Hsiang is supported by the National Science and Technology Council of Taiwan, R.O.C. under Grant No.~NSTC 113-2112-M-011-001-MY3.  B.-L. Hu thanks R. O. Behunin, his co-author in \cite{BehHu10,BH11}, for the highly valued work in establishing the graded influence action formalism for multi-partite open quantum systems. The present work began in 2013-2014 when both authors visited the physics department of Fudan University, Shanghai, China. They are thankful for the warm reception of Professor Y. S. Wu, then the Director of the new Fudan Center for Theoretical Physics. In 2023-24 B.-L. Hu enjoyed the  gracious hospitality of colleagues at the Institute of Physics, Academia Sinica and at the National Center for Theoretical Sciences at Tsing Hua University, Hsinchu, Taiwan, R.O.C. where this work resumed.

\end{document}